\documentclass[10pt]{bmc_article}

\usepackage{cite} % Make references as [1-4], not [1,2,3,4]
\usepackage{url}  % Formatting web addresses  
\usepackage{ifthen}  % Conditional 
\usepackage{multicol}   %Columns
\usepackage[utf8]{inputenc} %unicode support
\usepackage{graphicx}
\usepackage{color}

\usepackage{ulem}

\newboolean{publ}
\graphicspath{{graphics/}}

\newenvironment{bmcformat}{\begin{raggedright}\baselineskip20pt\sloppy\setboolean{publ}{false}}{\end{raggedright}\baselineskip20pt\sloppy}

\begin{document}
\begin{bmcformat}


\title{The Circuit Quantum Electrodynamical Josephson Interferometer}
 
%%%%%%%%%%%%%%%%%%%%%%%%%%%%%%%%%%%%%%%%%%%%%%
%%                                          %%
%% Enter the authors here                   %%
%%                                          %%
%% Ensure \and is entered between all but   %%
%% the last two authors. This will be       %%
%% replaced by a comma in the final article %%
%%                                          %%
%% Ensure there are no trailing spaces at   %% 
%% the ends of the lines                    %%     	
%%                                          %%
%%%%%%%%%%%%%%%%%%%%%%%%%%%%%%%%%%%%%%%%%%%%%%

\author{Robert Jirschik$^{1,2}$%
         \email{Robert Jirschik - r.jirschik.13@ucl.ac.uk}
       and 
         Michael J. Hartmann\correspondingauthor$^{1}$%
         \email{Michael J. Hartmann\correspondingauthor - mh@tum.de}%
      }

%%%%%%%%%%%%%%%%%%%%%%%%%%%%%%%%%%%%%%%%%%%%%%
%%                                          %%
%% Enter the authors' addresses here        %%
%%                                          %%
%%%%%%%%%%%%%%%%%%%%%%%%%%%%%%%%%%%%%%%%%%%%%%

\address{%
    \iid(1)Technische Universitaet Muenchen, Physik Department, James Franck Str., 85748 Garching, Germany\\
    \iid(2)Department of Physics and Astronomy, University College London, Gower Street, London WC1E 6BT, United Kingdom
}%

\maketitle

%%%%%%%%%%%%%%%%%%%%%%%%%%%%%%%%%%%%%%%%%%%%%%
%%                                          %%
%% The Abstract begins here                 %%
%%                                          %%  
%% Please refer to the Instructions for     %%
%% authors on http://www.biomedcentral.com  %%
%% and include the section headings         %%
%% accordingly for your article type.       %%   
%%                                          %%
%%%%%%%%%%%%%%%%%%%%%%%%%%%%%%%%%%%%%%%%%%%%%%

\begin{abstract}
Arrays of circuit cavities offer fascinating perspectives for exploring quantum many-body systems in a driven dissipative regime where excitation losses are continuously compensated by coherent input drives. Here we investigate a system consisting of three transmission line resonators, where the two outer ones are driven by coherent input sources and the central resonator interacts with a superconducting qubit.
Whereas a low excitation number regime of such a device has been considered previously with a numerical integration, we here specifically address the high excitation density regime.
We present analytical approximations to these regimes in the form of two methods.
The first method is a Bogoliubov or linear expansion in quantum fluctuations which can be understood as an approximation for weak nonlinearities.
As the second method we introduce a combination of mean-field decoupling for the photon tunneling with an exact approach to a driven Kerr nonlinearity which can be understood as an approximation for low tunneling rates.
In contrast to the low excitation regime we find that for high excitation numbers the anti-bunching of output photons from the central cavity does not monotonously disappear as the tunnel coupling between the resonators is increased. 
\end{abstract}

\ifthenelse{\boolean{publ}}{\begin{multicols}{2}}{}

%%%%%%%%%%%%%%%%%%%%%%%%%%%%%%%%%%%%%%%%%%%%%%
%%                                          %%
%% The Main Body begins here                %%
%%                                          %%
%% Please refer to the instructions for     %%
%% authors on:                              %%
%% http://www.biomedcentral.com/info/authors%%
%% and include the section headings         %%
%% accordingly for your article type.       %% 
%%                                          %%
%% See the Results and Discussion section   %%
%% for details on how to create sub-sections%%
%%                                          %%
%% use \cite{...} to cite references        %%
%%  \cite{koon} and                         %%
%%  \cite{oreg,khar,zvai,xjon,schn,pond}    %%
%%  \nocite{smith,marg,hunn,advi,koha,mouse}%%
%%                                          %%
%%%%%%%%%%%%%%%%%%%%%%%%%%%%%%%%%%%%%%%%%%%%%%

%%%%%%%%%%%%%%%%
%% Background %%
%%
\section{Introduction}
 
In recent years, a new direction of research in cavity quantum electrodynamics
(cavity-QED) has developed, in which multiple cavities that are
coupled via the exchange of photons are considered. Such setups are
particularly intriguing if the cavities are connected to form regular arrays
and if the strong coupling regime is achieved in each cavity of the
array. Such devices would give rise to novel structures where coherent light matter interactions
exceed dissipative processes simultaneously in multiple locations of the array
\cite{hartmann2006,angelakis2007,greentree2006,hartmann2008,tomadin2010,houck2012}.

Whereas it is rather challenging to build
mutually resonant high finesse cavities in the optical regime, it is for microwave photons
perfectly feasible to engineer large arrays of mutually
resonant cavities on one chip in an architecture known as
circuit-QED~\cite{houck2012,Lucero2012}.
Here multiple superconducting transmission line resonators with virtually identical lengths in the
centimeter range can be coupled via capacitors or inductive links \cite{houck2012}.
The individual transmission line resonators can feature strong optical nonlinearities by
coupling them to superconducting qubits such as transmons \cite{koch2007,fedorov2011} or phase qubits \cite{Lucero2012}.

Yet in all experiments that involve light-matter interactions, some photons will
inevitably be lost from the structure due to imperfect light
confinement or emitter relaxation. To compensate for such
losses, cavity arrays are thus most naturally studied in a
regime where an input drive continuously replaces the
dissipated excitations. This mode of operation eventually gives rise
to a driven dissipative regime, where the dynamical balance of loading
and loss processes leads to the emergence of stationary states
\cite{carusotto2009,nissen2012,Hartmann10,Jin13}. The
properties of these stationary states may however vary significantly if
one changes system parameters such as the photon tunneling rate between cavities,
the light-matter interaction in a cavity or even the relative phase between a pair of
coherent input drives \cite{gerace2009quantum,Hartmann10}.

A device that is ideally suited for studying the effect of relative phases between input drives is the so called quantum optical Josephson interferometer \cite{gerace2009quantum}, which consists of two coherently driven linear cavities connected through a central cavity with a single-photon nonlinearity. 
Here, the interplay between tunneling and interactions in the steady state of the system has been analyzed for regimes of weak input drives where the number of excitations in each cavity is rather small by Gerace et al. \cite{gerace2009quantum}. For opposite phases of the input drives one finds a destructive interference which suppresses population of the central cavity, whereas for input drives in phase the central cavity is populated with anti-bunched excitations due to its strong nonlinearity. 

For the considered regime of low excitation numbers the approach \cite{gerace2009quantum} employed a full numerical analysis relying on an excitation number truncation of the Hilbert spaces. For high input drives however, the theoretical description of the system poses a challenge as conventional numerical methods quickly become computationally infeasible. Hence, such a numerical approach cannot describe regimes with more intense input drives where excitation numbers begin to grow and is therefore unable to explore a possible transition from a quantum regime with anti-bunched output photons to a classical regime with uncorrelated output photons.

Here we present approaches that are capable of describing this transition. For a parameter regime with a weak nonlinearity in the central resonator we expand the intra cavity fields to linear order in the quantum fluctuations around their coherent parts. This Bogoliubov-type expansion provides a good approximation provided the density of quantum fluctuations around the coherent background is small compared to the excitation number in the latter. This regime is realized for weak nonlinearities in the central resonator. 

In turn for strong nonlinearities but moderate photon tunneling rates between the resonators we combine an exact solution for a driven Kerr nonlinearity by Drummond and Walls \cite{DrummondWalls} with a mean-field decoupling of the photon tunneling between resonators. As the solution is based on a description of the quantum state in terms of a P-function we will refer to it as P-function mean-field \cite{Boite13}. We find that this approach provides an accurate description in a large parameter range provided the photon tunneling is low enough such that the fields in the outer resonators do not deviate significantly from coherent ones. An overview of the discussed approaches and the parameter regimes where they provide an accurate description is sketched in figure \ref{regions}.
\begin{figure}
\centering
\includegraphics[width=0.5\textwidth]{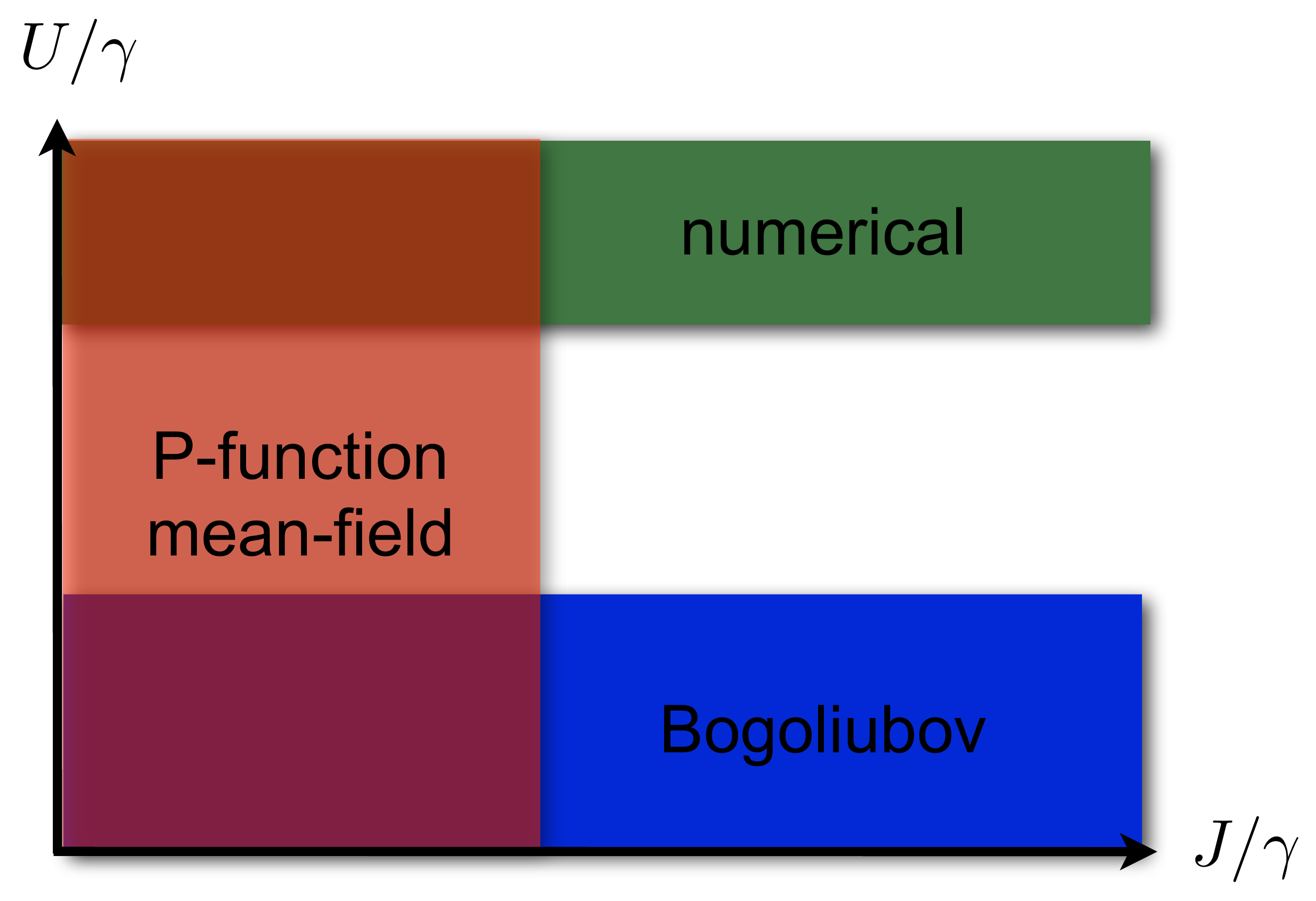}
\caption{Sketch of parameter regions where the discussed approaches provide accurate descriptions. With increasing driving strength $\Omega$, the validity range of the numerical approach shrinks to larger nonlinearities whereas the validity ranges for the Bogoliubov and P-function mean-field approaches grow. For definitions of $U$, $J$, $\Omega$ and $\gamma$ see equation (\ref{HSys}). The validity boundaries of the various approaches are discussed further in figures \ref{figure19}, \ref{figure20} and \ref{figure23a}.}
\label{regions}
\end{figure}

The remainder of the paper is organized as follows, in section \ref{sec:setup} we describe the setup and model we consider and
in section \ref{numericalsection} we revisit the numerical approach by Gerace et al. \cite{gerace2009quantum} to present results for time resolved correlation functions which so far have not been considered. In the next section \ref{quantum fluctuations} we discuss the approach based on a Bogoliubov expansion and the results it yields. In section \ref{analytical} we then introduce our mean-field approach based on an exact solution for the central resonator. Finally section \ref{sec:conclusions} presents conclusions and a discussion of the parameter regimes covered by each of the discussed descriptions.

\section{Setup} 
\label{sec:setup}
\begin{figure}
\centering
\includegraphics[width=0.45\textwidth]{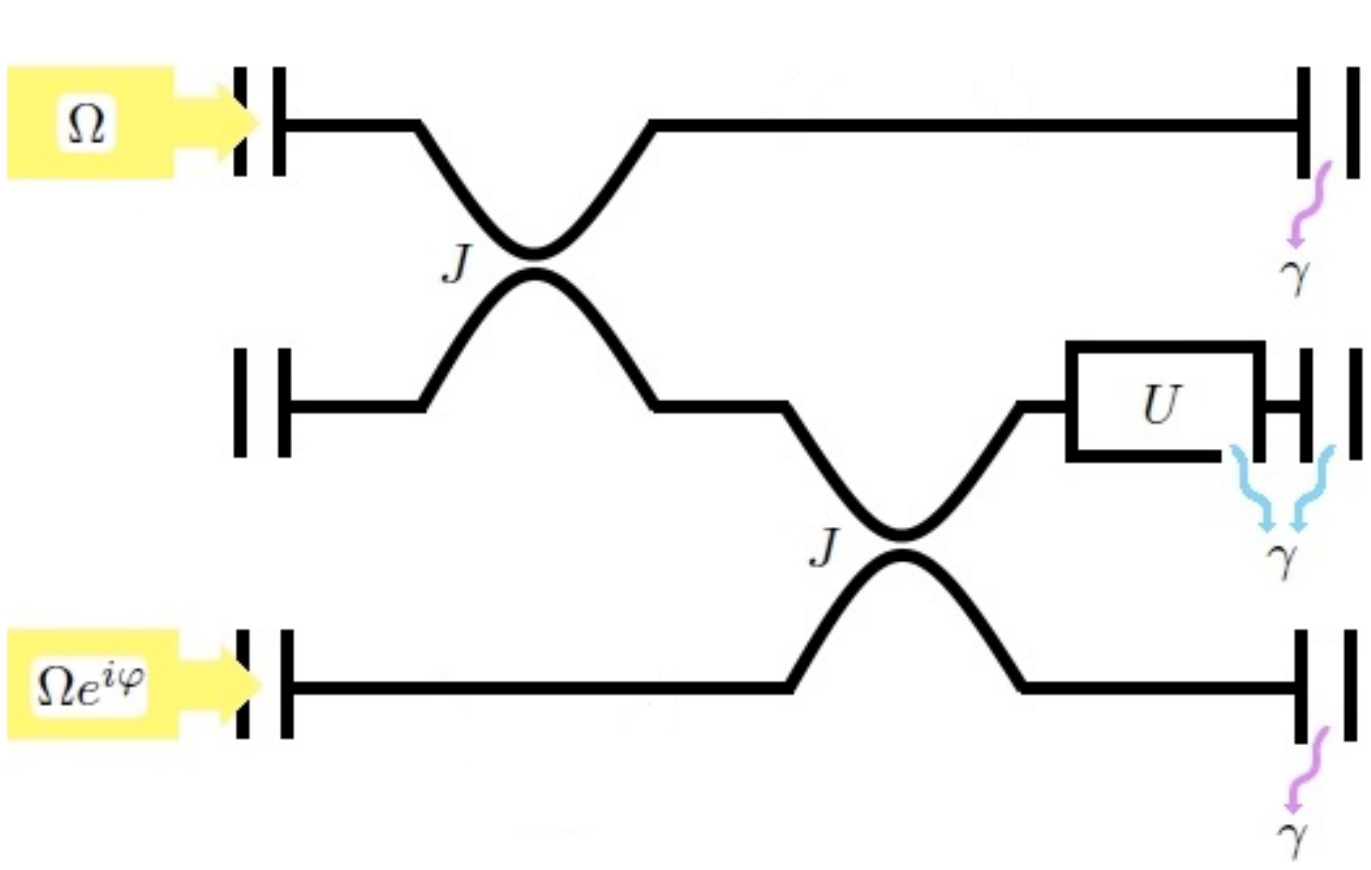}
\caption{Schematic of a possible realization of the quantum-optical Josephson interferometer with transmission line resonators in circuit QED. Here only the central conductors of the resonators are drawn. The three-resonator chip allows for a coupling $J$ between two neighboring resonators \cite{Peropadre13}. A voltage-node in the center is the reason for an off-center implementation of the nonlinearity $U$ that is capacitively coupled to the middle resonator. The two outer transmission line resonators are driven by coherent input sources with amplitudes $\Omega$. Each transmission line resonator has an individual dissipation-rate, which depends on the resonator's coupling capacitors at both ends. We consider the case where the combined decay rate of the central resonator and the transmon is equal to the decay rate of the two outer resonators $\gamma$.}
\label{figure1}
\end{figure}
The circuit QED Josephson interferometer we consider is a multicavity driven-dissipative system consisting of three coupled transmission line resonators where the two outer ones are coherently driven and the central transmission line resonator interacts with a superconducting qubit, see figure \ref{figure1} for a sketch. We consider here a transmon as the qubit and assume a regime with strong interactions between the central resonator and this qubit so that polaritons, combined light matter excitations, become a useful description of the nonlinear central resonator. For large enough qubit-resonator coupling one can focus the description only on one excitation species in the central resonator \cite{leib2010bose}. Employing this approach, the Hamiltonian for the circuit QED Josephson interferometer can be written in a frame rotating at the frequency $\omega_{in}$ of the input drives as
\begin{equation}\label{HSys}
\hat{H}=\sum^3_{k=1}\Delta_k\;\hat{a}^{\dagger}_k\hat{a}_k-\frac{U}{2}\hat{a}^{\dagger}_2\hat{a}^{\dagger}_2\hat{a}_2\hat{a}_2-J\left(\hat{a}_1^{\dagger}\hat{a}_2+\hat{a}_1\hat{a}_2^{\dagger}+\hat{a}_2^{\dagger}\hat{a}_3+\hat{a}_3^{\dagger}\hat{a}_2\right)+\frac{\Omega}{2}\left[\hat{a}_1^{\dagger}+\hat{a}_1+\hat{a}_3^{\dagger}e^{i\varphi}+\hat{a}_3e^{-i\varphi}\right].
\end{equation}
Here $\hat{a}_k$ annihilates an excitation in transmission line resonator $k$. For the two outer resonators the excitations are photons whereas for the central resonators they are excitations of the considered polariton mode. 
$\Delta_k = \omega_{k} - \omega_{in}$ is the detuning between the input drives and resonance frequency of excitations in the $k$-th transmission line resonator. For simplicity, we only consider the on-resonance case with $\Delta_k=0$. The parameter $U$ describes the effect of the Kerr-nonlinearity due to the transmon in the central resonator and $J$ is the tunneling rate between neighboring resonator sites, $\Omega$ is the drive amplitude, assumed to be equal for the two inputs, and $\varphi$ describes the phase difference between them.
At this point it is important to note that the nonlinearity of a transmon qubit is attractive (corresponding to $U>0$ in our notation).

The circuit QED Josephson interferometer is an open quantum system due to experimental imperfections such as the finite reflectivity of the coupling capacitors at each end of the transmission line resonators and relaxation of the transmon. We consider here the case where excitations in all three transmission line resonators have the same dissipation rates. 
As excitations in the second resonator can also decay via transmon relaxation, equal dissipation rates can be achieved for example by choosing suitable capacitances at the ends of the transmission lines. Since the couplings to the environments are sufficiently small \cite{fink2008climbing,bishop2008nonlinear} we can make use of the master equation approach \cite{carmichael1993open}. As a simplification we assume the reservoir state to be in vacuum, which is justified by the fact that circuit QED experiments are typically performed at a temperature in the millikelvin range \cite{wallraff2004strong}. These considerations lead to the master equation
\begin{equation}
\label{masterequ}
\dot{\rho}=\mathcal{L}\rho,
\end{equation}
where the Liouvillian $\mathcal{L}$ is given by
\begin{equation}
\mathcal{L}\rho=i\left[\rho,\hat{H}\right]+\frac{\gamma}{2}\sum^3_{k=1}\left(2\hat{a}_k\rho\hat{a}^{\dagger}_k-\hat{a}^{\dagger}_k\hat{a}_k\rho-\rho\hat{a}^{\dagger}_k\hat{a}_k\right),
\end{equation}
the dot denotes a time derivative and $\gamma$ describes the excitation loss rate of an individual resonator site.
Since the excitation losses are continuously compensated by a coherent input drive, a dynamical equilibrium leading to a steady state will be established. We are here interested in properties of this steady state, such as the mean excitation number and its correlation functions, mainly for the central resonator.

In cavity or circuit quantum electrodynamics experiments the output fields of the investigated resonators are typically accessible by measurements. Hence in our device the quantity of interest in measurements is the output field of the central resonator. It is linked to the field in the resonator via the input-output relation \cite{Collett84},
\begin{equation}
\label{inout1}
a_{out}(t) = \sqrt{\kappa} \, a_{res}(t) + a_{in}(t),
\end{equation}
where $a_{out}$ and $a_{in}$ are the output and input fields and $a_{res}$ the field in the resonator which leaks out at a rate $\kappa$.
In cavity and circuit QED the photons in the cavity usually couple to the photon modes in the output channel so that one can identify the cavity field $a_{res}$ with the annihilation operator of photons in the cavity. In the central cavity of our setup however, two polaritonic excitations build up due to the strong coupling to the qubit and the photon field in the central resonator can be expressed as a superposition of them, $a_{res} = \eta a_{2} + \eta' \tilde{a}_{2}$ with coefficients $\eta$ and $\eta'$ (Note that $\eta$ can always be chosen to be positive and that $|\eta|^{2} + |\eta'|^{2} = 1$). In the regime of strong coupling we are interested in, the frequencies of the polariton modes $a_{2}$ and $\tilde{a}_{2}$ differ strongly enough so that the input drives selectively only excite $a_{2}$ and we can neglect the $\tilde{a}_{2}$ excitations in the input-output relation since they only contribute vacuum output. Hence for calculating the output fields of the central resonator we can use the modified input-output relation,
\begin{equation}
\label{inout}
a_{out}(t) = \sqrt{\gamma} \, a_{2}(t) + a_{in}(t),
\end{equation}
where $\gamma = \eta \kappa$. For vacuum input fields, we thus obtain an out photon flux $\langle a_{out}^{\dagger} a_{out}\rangle = \gamma \langle a_{2}^{\dagger} a_{2} \rangle$ for the central resonator.

The quantumness or non-classicality of light fields can be characterized by the statistics of their excitations. Here a quantity of central interest is
the so-called (time-resolved) second order correlation function, here for resonator 2,
\begin{equation}
\label{g2tau}
g^{(2)}(\tau)=\frac{\langle\hat{a}_{2}(t)^{\dagger}\hat{a}_{2}^{\dagger}(t+\tau)\hat{a}_{2}(t+\tau)\hat{a}_{2}(t)\rangle}{|\langle\hat{a}_{2}(t)^{\dagger}\hat{a}_{2}(t)\rangle|^2}
\end{equation}
The significance of the physical interpretation is to have a value for the probability to measure a second particle at time $t+\tau$ after a particle has been measured at time $t$ and compare this to a coherent field where $g^{(2)}=1$. Therefore, if $g^{(2)}<1$ it is less likely to measure a second particle. In this case we speak of an anti-bunched light field that is necessarily non-classical, whereas for $g^{(2)}>1$ we have bunched light meaning a higher probability than for a coherent field.
We note that the $g^{(2)}$-function for the output fields is identical to the $g^{(2)}$-function of intra-cavity fields provided the input fields are in vacuum.

Let us finally stress that although we refer to an implementation in circuit electrodynamics here, our calculations do in most parts not make use of details of this technology and are therefore applicable to other realizations as well. Yet we prefer to discuss them in the context of circuit QED as this technology offers excellent perspectives for realizing a coupling between resonators as we envision it here.

\section{The Numerical Approach in the Weak Driving Regime}\label{numericalsection}
In the regime where the laser intensity is rather low, we are able to numerically solve the master equation directly. Here, an entire regime of rather high nonlinearities $U/\gamma>1$, which lessen the probability of high intra-cavity excitations, can be covered for arbitrary values of the tunneling rate $J$. We introduce a cut-off in the excitation number to give a matrix expression for the excitation ladder operators in a Fock basis of bosonic number states. This reduces the infinitely high dimensional Hilbert space to a small number of dimensions and hence allows for the numerical calculation of all operator expectation-values. To corroborate our results we test their convergence as we increase the excitation number cut-off.

For the steady states we consider, $g^{(2)}(\tau)$ as given in equation (\ref{g2tau}) is independent of the time $t$ and only depends on the time delay $\tau$. In the weak driving regime the results for no delay $\tau = 0$ have already been covered in \cite{gerace2009quantum} and we therefore focus for this regime on the time-resolved second order correlation function for $\tau \ne 0$. 

Applying the quantum regression theorem (QRT) \cite{lax1963formal} we find the results presented in Fig. \ref{figure8}. Here, as expected, the $g^{(2)}(\tau)$-function converges to 1 for a long time delay $\tau\rightarrow\infty$. The amplitudes of the flucutations around $g^{(2)}(\tau)=1$ increase with increasing nonlinearities and increasing tunneling rates $J$, however at $\tau=0$ the system is more excitation anti-bunched if the tunneling strength is weaker and the nonlinearity higher.
\begin{figure}
\centering
\includegraphics[width=0.4\textwidth]{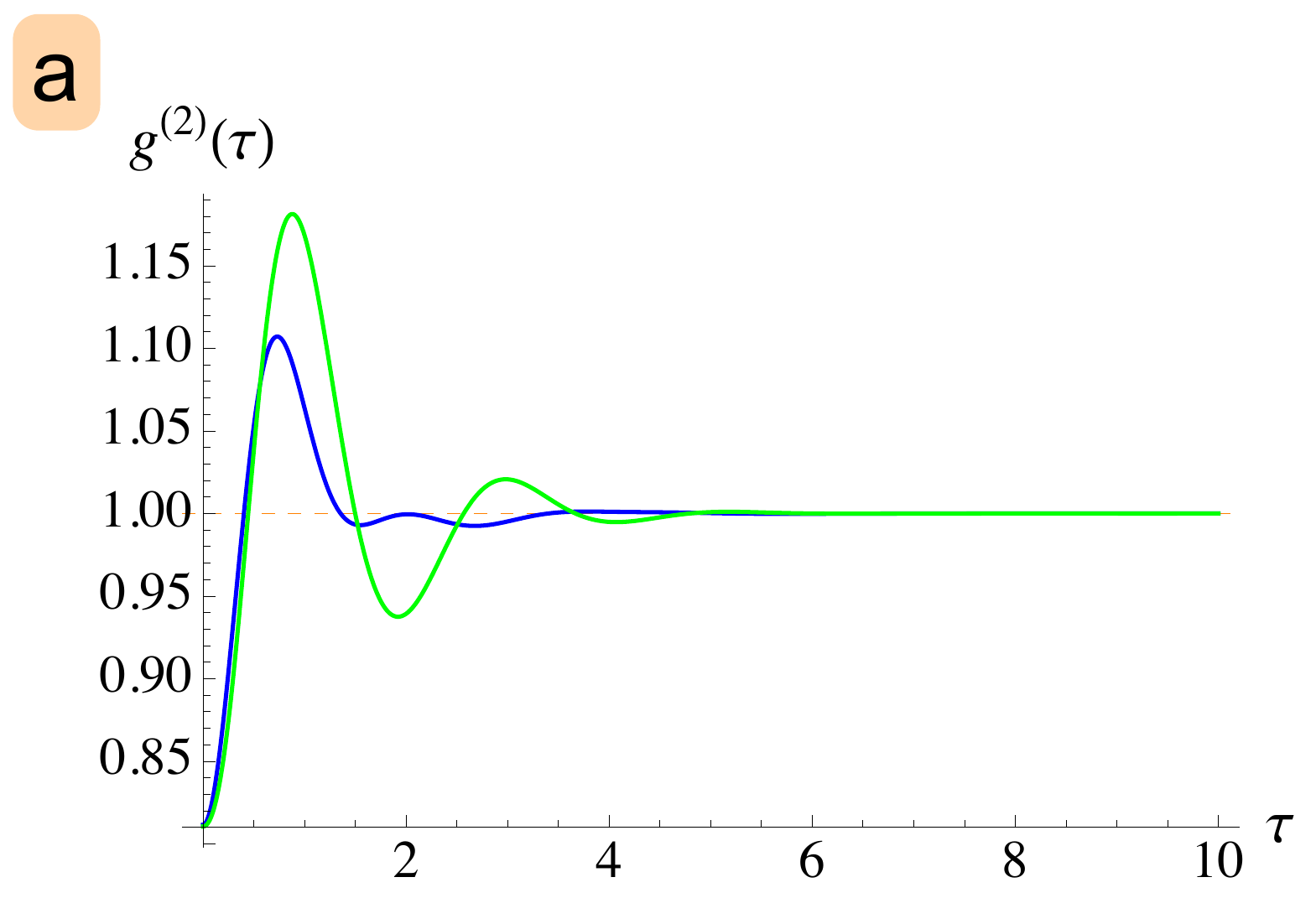}
\hspace{0.4cm}
\includegraphics[width=0.4\textwidth]{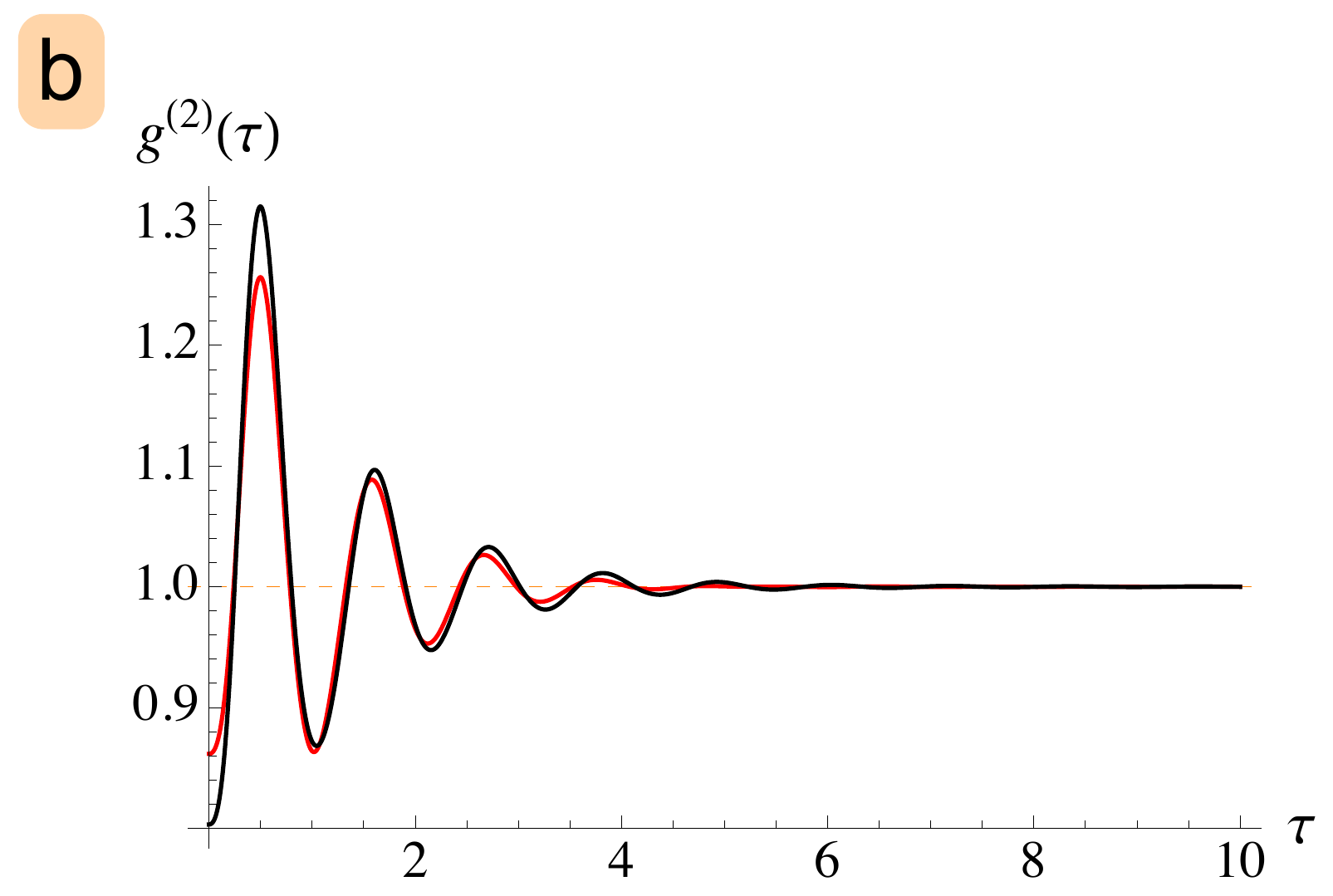}
\caption{The second order correlation function $g^{(2)}(\tau)$ as found from a fully numerical solution. {\bf a}: for $\varphi=\Delta=0$, $J/\gamma=2$, $U/\gamma=2$ and $\Omega/\gamma=1$ (green) or $\Omega/\gamma=2$ (blue). {\bf b}: for $\varphi=\Delta=0$, $J/\gamma=4$, $U/\gamma=2$ and $\Omega/\gamma=1$ (black) or $\Omega/\gamma=2$ (red).}
\label{figure8}
\end{figure}

With the used approach of directly solving the master-equation numerically in a truncated Hilbert-space one can cover a regime for arbitrary values of $J$ if the nonlinearity fulfills $U/\gamma>1$ in the weak driving regime of $|\Omega|/\gamma<2$. Nevertheless, this method is unable to show a clear crossover between anti-bunched and uncorrelated excitations because the weak drive and high nonlinearity restrict it to low numbers of excitations, implying anti-bunched statistics. Therefore we will present a method to extend the investigation of this set-up to the strongly driven regime, motivated by the possibility of high intra-resonator excitations.
%_________________________________________________________________________________________________________________________________________

\section{The Quantum Fluctuations Approach in the Strong Driving Regime}\label{quantum fluctuations}

For higher laser intensities, the number of excitations increases rapidly. Since the two outer transmission line resonators do not directly couple to qubits, their only deviation from a coherent state stems from the coupling to the central resonator, which features a nonlinearity due to its coupling to a transmon qubit. Therefore it is for strong inputs no longer computationally feasible to numerically solve the full master equation. Assuming a weak nonlinearity, we thus use a generalized Bogoliubov approach to non-equilibrium steady states \cite{Hartmann10}, which linearizes the equations for the quantum fluctuations around the classical fields and can be understood as an approximation for low nonlinearities. We expand the creation and annihilation operators around the coherent state by writing
\begin{eqnarray}
\label{expansion1}
\hat{a}_j&=&\alpha_j+\widehat{\delta a}_j\\
\label{expansion2}
\hat{a}^{\dagger}_j&=&\alpha^*_j+\widehat{\delta a}^{\dagger}_j
\end{eqnarray}
where the complex number $\alpha_j=\langle\hat{a}_j\rangle$ represents the coherent background and the operators $\widehat{\delta a}_j$ and $\widehat{\delta a}^{\dagger}_j$ describe the quantum fluctuations around it. 
Using equations (\ref{expansion1}) and (\ref{expansion2}) we expand the master equation in powers of the operators $\widehat{\delta a}_j$ and $\widehat{\delta a}^{\dagger}_j$, 
\begin{equation}
\dot{\rho}=\left(\mathcal{L}^{(1)}+\mathcal{L}^{(2)}+\mathcal{L}^{(3)}+\mathcal{L}^{(4)}\right)\rho
\end{equation}
where $\mathcal{L}^{(n)}\left[\rho\right]$ denotes the $n$th order of quantum fluctuations $\widehat{\delta a}_j$ in the Liouvillian $\mathcal{L}$.

\subsection{Steady State Coherent Background}\label{coherentbackground}
To determine the coherent background in terms of the $\alpha_{j}$, we require that the first order Liouvillian vanishes for all quantum fluctuations, $\mathcal{L}^{(1)}\stackrel{!}{=}0$. The result is equivalent to the solution for the steady state coherent background.
With the above requirement we get a cubic equation for the steady state mean excitation number $n_2=|\alpha_2|^2$ of the coherent background in the second cavity
\begin{equation}
4\gamma^2U^2n_2^3+\left(\gamma^2+8J^2\right)^{2}n_2=8J^2\Omega^2\left(1+\cos{\varphi}\right)
\end{equation}
which has only one physically meaningful solution with $n_2\geq0$:
\begin{equation}
\label{excitationnumber}
n_2=\frac{\eta^2-3^{1/3}U^2\gamma^2\left(\gamma^2+8J^2\right)^2}{2\times3^{2/3}U^2\gamma^2\eta}
\end{equation} 
where
\begin{equation}
\eta=\sqrt[3]{72 \gamma ^4 J^2 U^4 \Omega ^2 (\cos (\varphi )+1)+\sqrt{3} \sqrt{\gamma ^6 U^6 \left(6912 \gamma ^2 J^4 U^2 \Omega ^4 \cos ^4\left(\frac{\varphi }{2}\right)+\left(\gamma ^2+8 J^2\right)^6\right)}}
\end{equation}
The values of $\alpha_{1}$ and $\alpha_{3}$ can be determined from the solution (\ref{excitationnumber}) via the additional relations
\begin{eqnarray}
\label{alpha1}\alpha_1&=&\frac{i\left(2J\alpha_2-\Omega\right)}{\gamma},\\
\label{alpha3}\alpha_3&=&\frac{i\left(2J\alpha_2-\Omega e^{i\varphi}\right)}{\gamma}.
\end{eqnarray}

The solution (\ref{excitationnumber}) is plotted in Fig. \ref{figure11}.
One can clearly see that the coherent background decreases for increasing nonlinearities. This feature appears since the frequencies of the input drives are resonant with the transition between zero- and single-excitation states in the resonators. Hence with increasing nonlinearity it becomes less probable that higher excitation states become populated as well. Moreover the population of the coherent part of the intra-resonator fields shows a maximum for specific values of
$J\geq0$. These maxima can be understood when writing the Hamiltonian (\ref{HSys}) in terms of Bloch modes \cite{Hartmann10}, which have frequencies that depend on the tunneling rate $J$. Thus with increasing $J$ a balance establishes between the higher detuning between Bloch modes and input drives leading to less excitations and an increased tunneling leading to more excitations in the central cavity. Here, it can already be seen that the number of polaritons in the second resonator exceeds the computational feasibility of conventional numerical methods to calculate the 3-site system.
\begin{figure}
\centering
\includegraphics[width=0.4\textwidth]{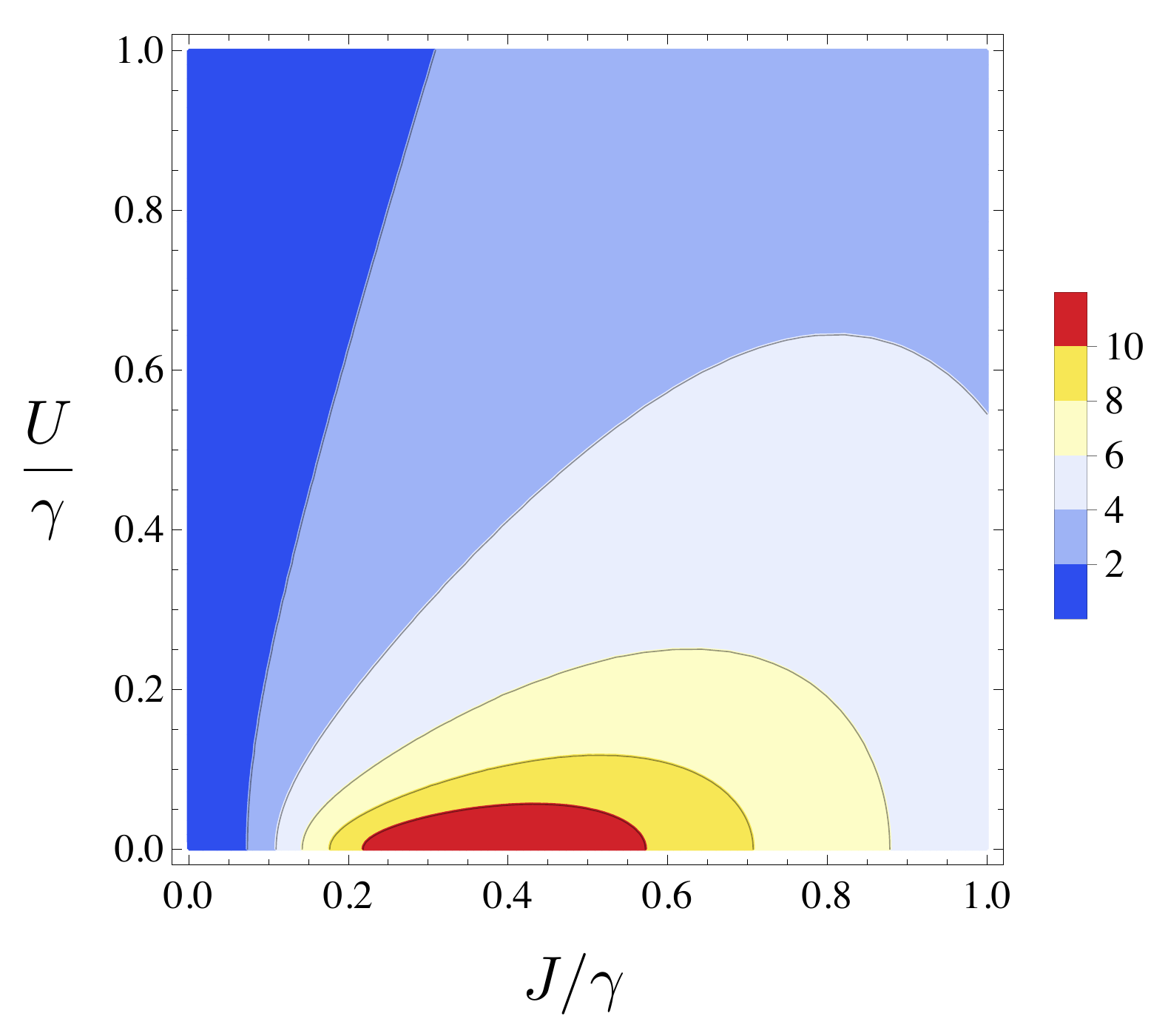}
\caption{The coherent excitation number $n_{2}=|\alpha_2|^2$ in the second transmission line resonator as a function of the nonlinearity $U/\gamma$ and the tunneling rate $J/\gamma$ for in-phase laser input $\varphi=0$ and $\Omega/\gamma=5$. Stronger drives lead to even higher excitation numbers $n_{2}$.}
\label{figure11}
\end{figure}

Fig. \ref{figure12} shows the effects of varying the phase-difference $\varphi$ of the two coherent input laser sources. One immediately notices the emptying of the central cavity for a phase-difference $\varphi = ( 2m +1) \pi$ with integer $m$. The oscillation of the excitation number in the central cavity is a result of the interference between the two coherent input drives. On the other hand, changing the tunneling-rate does not affect the minimum, but changes the severeness of the interference effect in agreement with the previously visualized results in Fig. \ref{figure11}. We further find (not plotted) that for an increase in driving strength the oscillation due to the phase-difference $\varphi$ is steeper, changes the value of $J$ for which the  background occupation number has a maximum and decreases the incline around it. Also the sharpness of the peak of maximal background occupation is less pronounced for higher nonlinearities.
\begin{figure}
\centering
\includegraphics[width=0.4\textwidth]{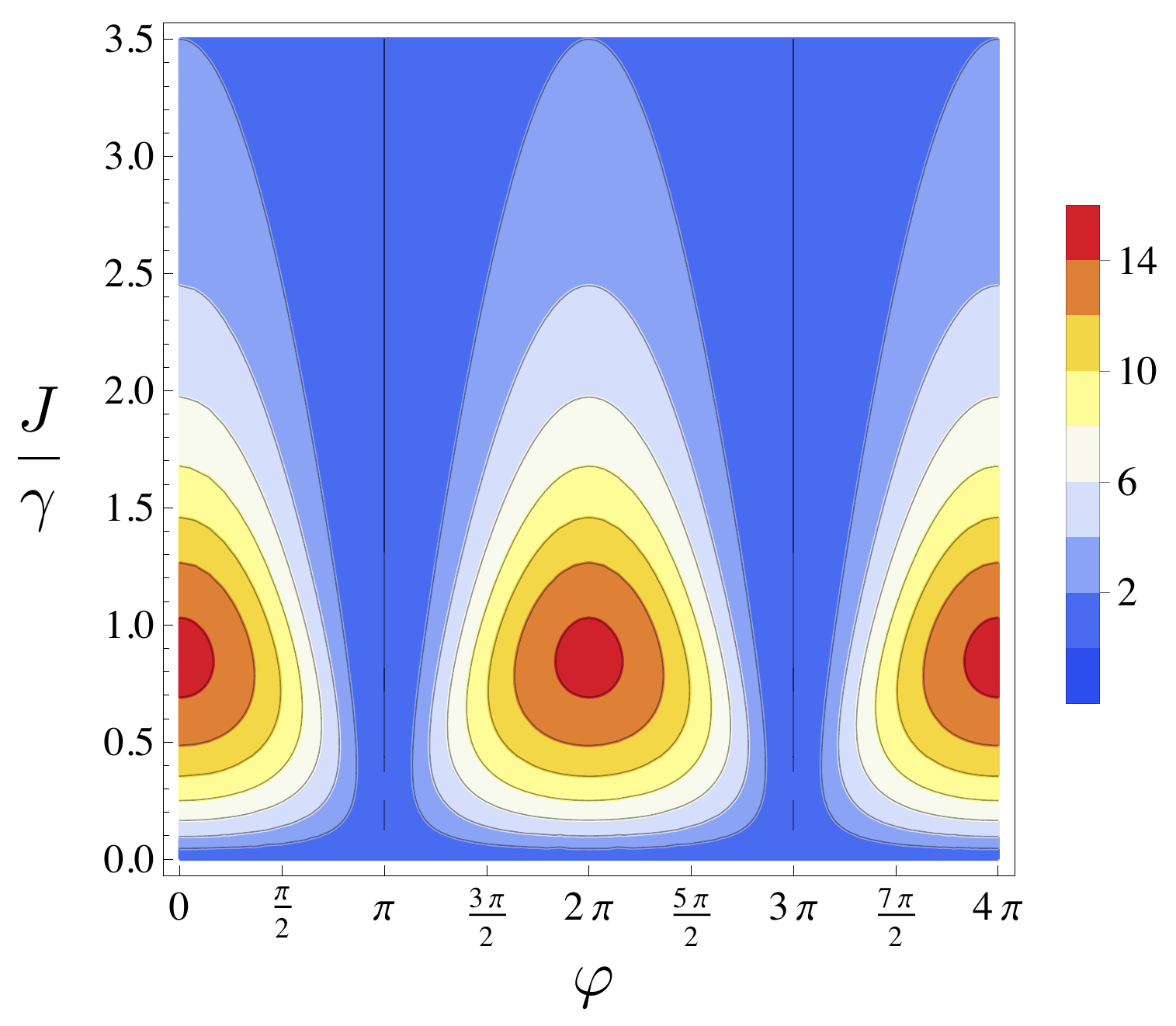}
\caption{Background excitation number $n_{2}$ as a function of the phase difference $\varphi$ and tunneling $J/\gamma$. Due to a destructive interference, excitations are suppressed for $\varphi = ( 2m +1) \pi$ with integer $m$. The plot shows the effect for a low nonlinearity $U/\gamma=0.2$ and a strong drive $\Omega/\gamma=10$.}
\label{figure12}
\end{figure}

%_______________________________________________________________________________________________________________________________________
\subsection{Steady State Quantum Fluctuations}
We are mainly interested in the statistics of output photons from the central resonator. These are described by the mean excitation number $\langle\hat{n}_2\rangle$ and the second-order correlation function $g^{(2)}_2$.
To calculate the mean excitation number in the steady state, we can write in the expansion approach,
\begin{equation}
\label{excitation_number}
\langle\hat{a}^{\dagger}\hat{a}\rangle=|\alpha|^2+\langle\widehat{\delta a}^{\dagger}\widehat{\delta a}\rangle,
\end{equation}
where we have dropped the site index as we are only focusing on resonator 2.
The second order correlation function has been introduced in Eq. (\ref{g2tau}). The case with zero time-delay $\tau=0$ is important in order to determine whether the output light field exhibits Poissonian or sub-Poissonian statistics. We expand the $g^{(2)}$-function to get
\begin{equation}\label{g2_expression}
g^{(2)}\left(\tau=0\right)=\frac{|\alpha|^4+4|\alpha|^2\langle\widehat{\delta a}^{\dagger}\widehat{\delta a}\rangle+\left(\alpha^*\right)^2\langle\widehat{\delta a}\widehat{\delta a}\rangle+\left(\alpha\right)^2\langle\widehat{\delta a}^{\dagger}\widehat{\delta a}^{\dagger}\rangle}{|\alpha|^4+2|\alpha|^2\langle\widehat{\delta a}^{\dagger}\widehat{\delta a}\rangle}.
\end{equation}
Therefore, in order to calculate the important statistical properties of our system we need to combine the previously derived results of the coherent background with the mean correlation values of the quantum fluctuations. 
We thus solve the master equation (\ref{masterequ}) to second order in the quantum fluctuations for $\rho$,
\begin{equation}\label{2ndorder}
\mathcal{L}^{(2)}\rho=0,
\end{equation}
using $\mathcal{L}^{(1)}=0$ and plugging the solution (\ref{excitationnumber}) into $\mathcal{L}^{(2)}$.
For the considered case of $\Delta = 0$, the eigenvalues of $\mathcal{L}^{(2)}$ all have real parts $-\gamma/2$ so that the solutions are stable.
Neglecting higher order terms in the master equation, i.e. $\mathcal{L}^{(3),(4)}\approx0$, is a good approximation provided that 
quantum fluctuations are small compared to the coherent background,
\begin{equation}
\label{ratio}
\langle\widehat{\delta a}_j^{\dagger}\widehat{\delta a}_j\rangle\ll|\alpha_j|^2 \quad \textrm{for} \: j = 1,2,3.
\end{equation} 
To analyze the validity of the solutions we find, we thus check whether they fulfill this property.

For the central transmission line resonator, the solution for the mean quantum fluctuation number reads
\begin{eqnarray}
\label{qadqa}\langle\widehat{\delta a}_2^{\dagger}\widehat{\delta a}_2\rangle &=& \frac{2 n_2^2 U^2 \left(\gamma ^4+16 J^4+3 \gamma ^2 \left(2 J^2+n_2^2 U^2\right)\right)}{\left(\gamma ^2+3 n_2^2 U^2\right) \left(\gamma ^4+64 J^4+4 \gamma ^2 \left(4 J^2+3 n_2^2 U^2\right)\right)}
\end{eqnarray}
Considering the fact that we neglected the terms of third and fourth order in quantum fluctuations, which contain additional pre-factors of the nonlinearity $U$, we come to the conclusion that our approach can also be considered an expansion in $U$.
As a consequence, we can deduce the results represent a good approximation in a regime of weak nonlinearities $U$. Moreover, the approach becomes much better for higher laser-input intensities because the coherent background excitation number in Eq.(\ref{excitationnumber}) scales much stronger with the laser input amplitude $\Omega$ than the number of quantum fluctuations in Eq.(\ref{qadqa}).

In contrast to the mean excitation number in the second resonator, the quantum fluctuations do play a significant role for the 
$g^{(2)}$-function, see Eq. (\ref{g2_expression}). 
The second order correlation function of the central resonator in the strong driving regime is plotted in Fig. \ref{figure17}. Depending on the tunneling rate $J$ we observe smooth transitions from a coherent to an anti-bunched excitation field as the nonlinearity is increased beyond specific values.
\begin{figure}
\centering
\includegraphics[width=0.4\textwidth]{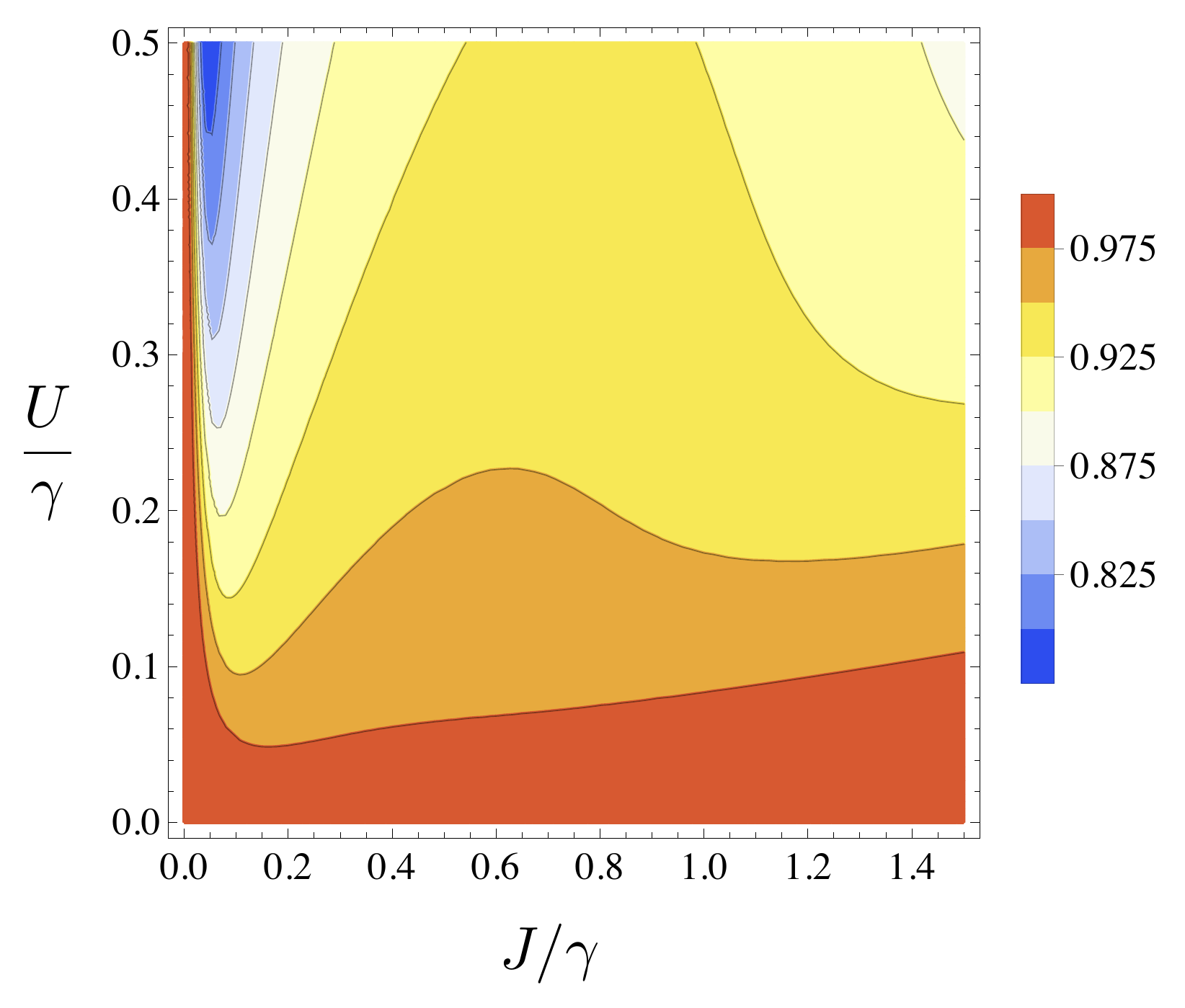}
\caption{The $g^{(2)}$-function as given by equation (\ref{g2_expression}) as a function of $J/\gamma$ and $U/\gamma$ in the in-phase strong driving regime for $\Omega/\gamma=5$. The red color indicates a coherent field, where $g^{(2)}\approx1$, whereas the blue area marks excitation-antibunching with $g^{(2)}<1$.}
\label{figure17}
\end{figure}
Fig. \ref{figure18} shows that the oscillations in the excitation number due to interference effects can be observed in the $g^{(2)}$-function as well. 
\begin{figure}
\centering
\includegraphics[width=0.4\textwidth]{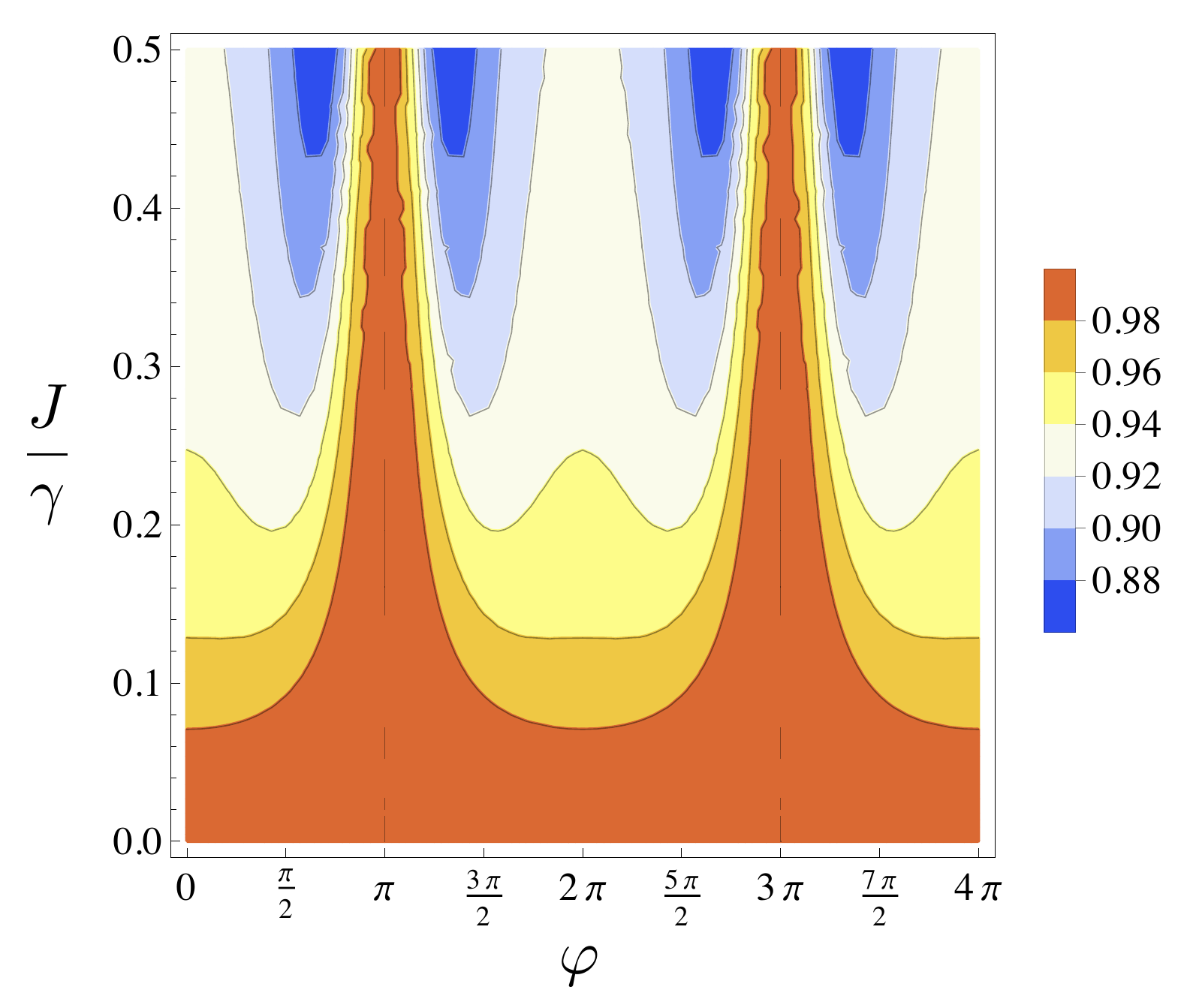}
\caption{The $g^{(2)}$-function as given by equation (\ref{g2_expression}) as a function of $\varphi$ and $U/\gamma$ for $J/\gamma=1$ and $\Omega/\gamma=5$. The red color indicates a coherent field, where $g^{(2)}\approx1$, whereas the blue area marks excitation-antibunching with $g^{(2)}<1$.}
\label{figure18}
\end{figure}

%_________________________________________________________________________________________________________________________________________
\subsection{Limitations of the Bogoliubov Approach}
The major limitation of the Bogoliubov approach is obviously the validity of the assumption (\ref{ratio}).
In order to determine the regime where this is fulfilled, we plot the ratio $\langle\widehat{\delta a}^{\dagger}_i\widehat{\delta a}_i\rangle /\left|\alpha_i\right|^2$ ($i=1,2,3$). For parameters where this ratio is sufficiently small, the approximation works fairly well.
The ratio $\langle\widehat{\delta a}^{\dagger}_2\widehat{\delta a}_2\rangle/\left|\alpha_2\right|^2$ is plotted in Fig. \ref{figure19}, whereas Fig. \ref{figure20} depicts the same ratio for the two outer resonator sites, i.e. $\langle\widehat{\delta a}^{\dagger}_1\widehat{\delta a}_1\rangle/\left|\alpha_1\right|^2 = \langle\widehat{\delta a}^{\dagger}_3\widehat{\delta a}_3\rangle/\left|\alpha_3\right|^2$.
\begin{figure}
\centering
\includegraphics[width=0.4\textwidth]{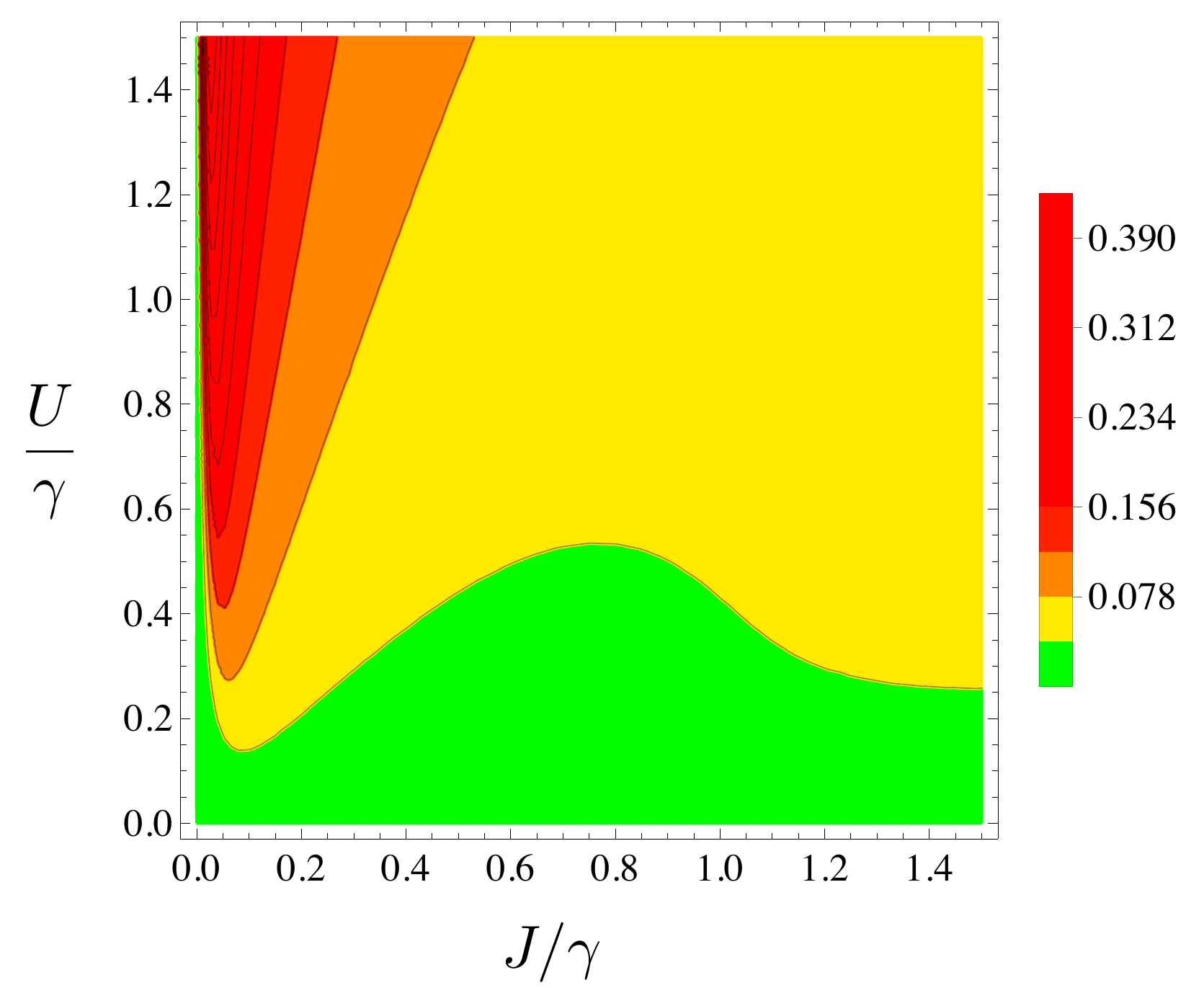}
\caption{Ratio of quantum fluctuations to the coherent background excitation number for the mid-resonator, $\langle\widehat{\delta a}^{\dagger}_2\widehat{\delta a}_2\rangle/\left|\alpha_2\right|^2$ as a function of $J/\gamma$ and $U/\gamma$ in the strong driving regime with $\Omega/\gamma=5$.
For higher drives $\Omega$ the ratio becomes smaller and also stays small up to larger nonlinearities $U$.}
\label{figure19}
\end{figure}
\begin{figure}
\centering
\includegraphics[width=0.4\textwidth]{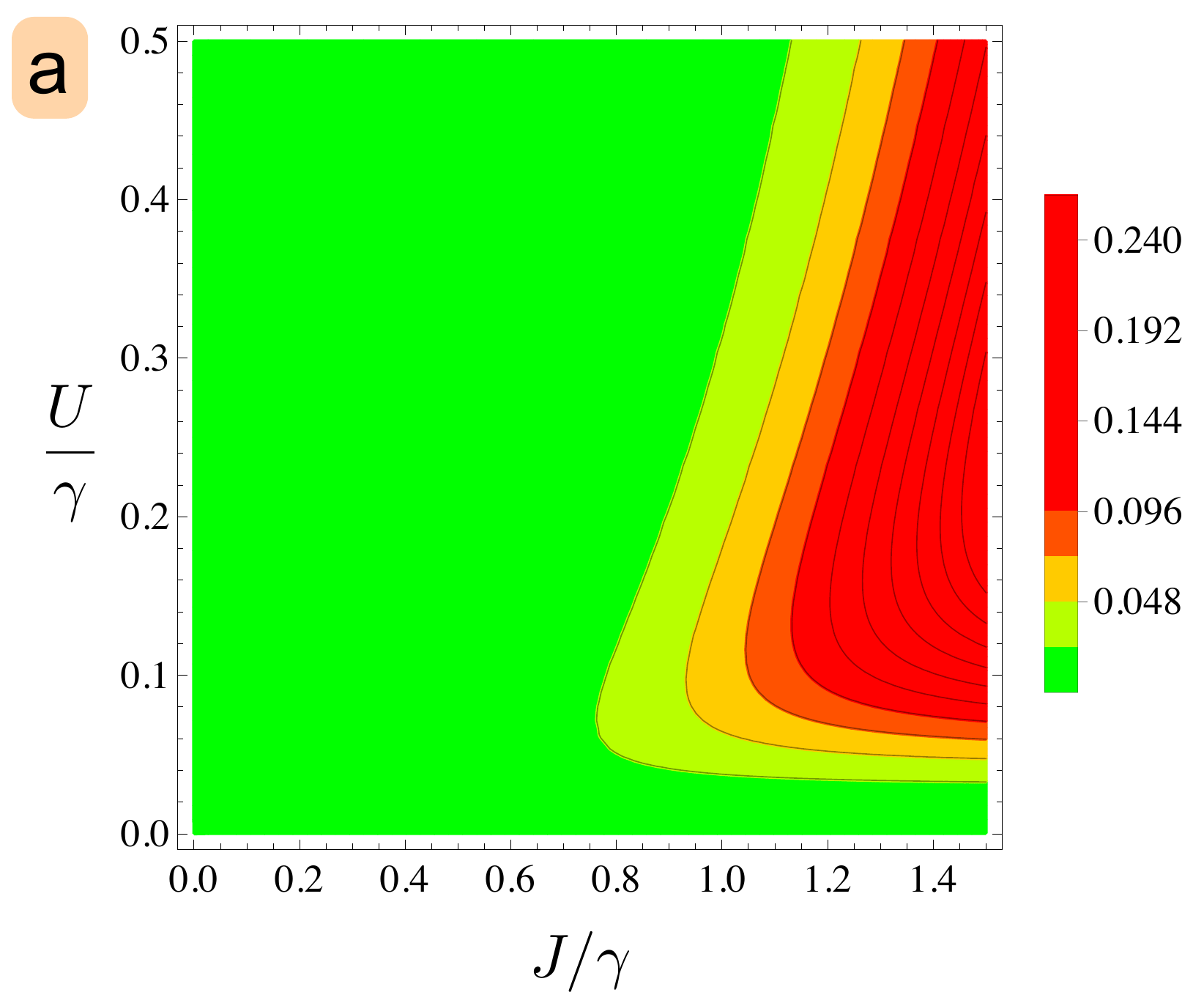}
\hspace{0.4cm}
\includegraphics[width=0.4\textwidth]{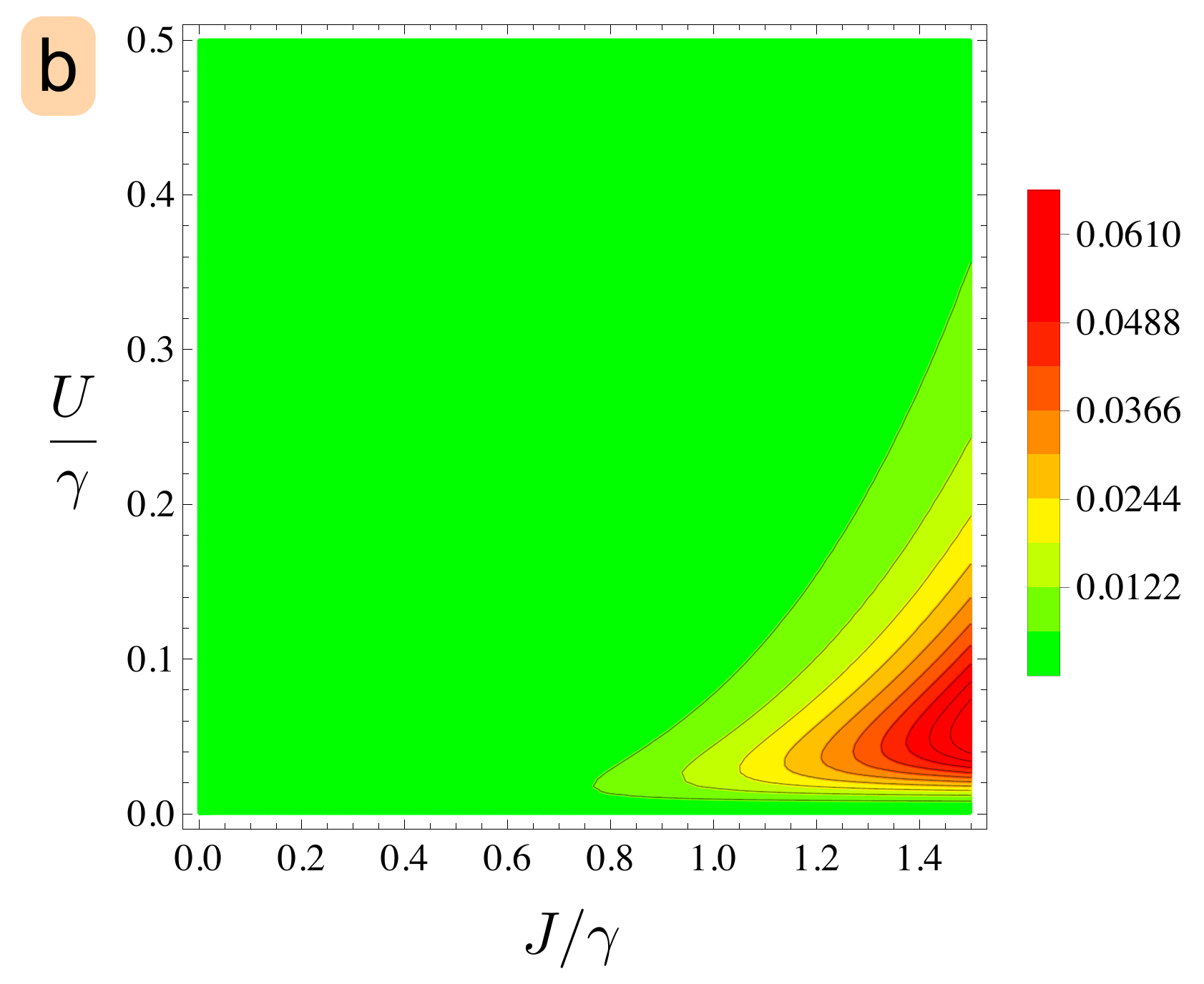}
\caption{Ratio of quantum fluctuations to the coherent background excitation number for the two outer resonators, $\langle\widehat{\delta a}^{\dagger}_1\widehat{\delta a}_1\rangle/\left|\alpha_1\right|^2 = \langle\widehat{\delta a}^{\dagger}_3\widehat{\delta a}_3\rangle/\left|\alpha_3\right|^2$ as a function of $J/\gamma$ and $U/\gamma$, {\bf a}: $\Omega/\gamma=5$ {\bf b}: $\Omega/\gamma=10$.}
\label{figure20}
\end{figure}

As a conclusion, we can safely assume that our approximation works well in a regime for strong input drives $\Omega/\gamma>5$ and low nonlinearities $U/\gamma<0.2$ for arbitrary values of the tunneling rate. For larger drives, such as $\Omega/\gamma>10$ the range of validity extends to larger $U/\gamma$. Thus we are motivated to find access to the regime of larger nonlinearities $U/\gamma$ because, as is evident from Fig. \ref{figure19}, the assumption $\langle\widehat{\delta a}_{i}^{\dagger}\widehat{\delta a}_{i}\rangle\ll|\alpha_{i}|^2$ breaks down in that regime, primarily in the second resonator. We present an approach to this parameter regime in the next section.

%_________________________________________________________________________________________________________________________________________

\section{P-Function Mean-Field Approach to the Low Tunnelling Regime}\label{analytical}
As one can see from figures \ref{figure19} and \ref{figure20}, the main limitation of a Bogoliubov approximation is the ratio of the mean quantum fluctuation number to the mean excitation number in the second cavity which significantly increases for large nonlinearities $U$ and weak drives $\Omega$. Hence if we are able to take higher quantum fluctuations in resonator 2 into account the results should become more exact and be applicable to a larger parameter range. On the other hand the deviation from a coherent state for the first and third transmission line resonator is extremely small because they do not contain any nonlinearity themselves and the nonlinearity of the center cavity only slightly affects the two outer cavities.
Thus, if we are able to find a way to solve the second transmission line resonator's mean values exactly, our results should become better. This will be the ansatz of our mean-field approach where we ignore quantum fluctuations in the two outer resonators and substitute $\hat{a}_{1,3}\rightarrow\alpha_{1,3}$. With this substitution, the Hamiltonian for resonator 2 takes the form
\begin{equation}
\hat{H}_{MF}=\Delta\omega\hat{a}_2^{\dagger}\hat{a}_2-\frac{U}{2}\hat{a}_2^{\dagger}\hat{a}_2^{\dagger}\hat{a}_2\hat{a}_2-J\left(\alpha_1+\alpha_3\right)\hat{a}_2^{\dagger}-J\left(\alpha^*_1+\alpha^*_3\right)\hat{a}_2
\end{equation}
which can be written in the form of a coherently driven Kerr nonlinearity \cite{DrummondWalls},
\begin{equation}\label{HMF}
\hat{H}_{MF}=\Delta\omega\hat{a}_2^{\dagger}\hat{a}_2-\frac{U}{2}\hat{a}_2^{\dagger}\hat{a}_2^{\dagger}\hat{a}_2\hat{a}_2+\frac{\tilde{\Omega}}{2}\hat{a}_2^{\dagger}+\frac{\tilde{\Omega}^*}{2}\hat{a}_2
\end{equation}
by introducing the drive $\tilde{\Omega}=-2J\left(\alpha_1+\alpha_3\right)$. To determine $\alpha_1$ and $\alpha_3$ self consistently, we use equations (\ref{alpha1}) and (\ref{alpha3}) to obtain,
\begin{equation}
\label{newdrive}
\tilde{\Omega}=\frac{2iJ}{\gamma}\left(\Omega+\Omega e^{i\varphi}-4J\langle\hat{a}_2\rangle\right)
\end{equation}
in the $\Delta=0$ case.
Our approach can thus be understood as a mean-field decoupling of the three cavities. With the consistency condition (\ref{newdrive}),
the Hamiltonian (\ref{HMF}) becomes a single site model which can be solved exactly using a P-function based method introduced by Drummond and Walls \cite{DrummondWalls}.

What we have done is basically a trade-off: We allow for a more crude approximation of the quantum tunneling effects in the two outer cavities by completely neglecting their quantum fluctuations. The latter are however expected to be very small, see Fig. \ref{figure20}. In turn we gain an exact solution for the resulting one-site problem, giving a better approximation of the nonlinear effects in the second transmission line resonator. In contrast to the Bogoliubov expansion that has been performed in section~\ref{quantum fluctuations}, we here do not expand in $U$ but in the tunneling rate $J$ \cite{degenfeld2013self}. With the approach introduced here we thus cover a regime with arbitrary values for the nonlinearity $U$ but only for moderate values of $J$. 
The solution for the normal-ordered mean values of the middle transmission line resonator in the $\Delta=0$ case reads
\begin{equation}\label{exactsolution}
\langle\left(\hat{a}^{\dagger}_2\right)^k\left(\hat{a}_2\right)^l\rangle=\left(\frac{\tilde{\Omega}^*}{U}\right)^k\left(\frac{\tilde{\Omega}}{U}\right)^l\frac{\Gamma(-i\frac{\gamma}{U})\Gamma(i\frac{\gamma}{U})}{\Gamma(k-i\frac{\gamma}{U})\Gamma(l+i\frac{\gamma}{U})}\frac{_0\mathcal{F}_2(k-i\frac{\gamma}{U},l+i\frac{\gamma}{U},2|\frac{\tilde{\Omega}}{U}|^2)}{_0\mathcal{F}_2(-i\frac{\gamma}{U},i\frac{\gamma}{U},2|\frac{\tilde{\Omega}}{U}|^2)},
\end{equation}
where 
\begin{equation}
_0\mathcal{F}_2(c,d,z)=\sum^{\infty}_{n=0}\left(\frac{z^n\Gamma(c)\Gamma(d)}{\Gamma(c+n)\Gamma(d+n)n!}\right)
\end{equation}
is the generalized Gauss hypergeometric series, defined in terms of the $\Gamma$-function. Note that $\tilde{\Omega}$ is a function of
$\langle\hat{a}_2\rangle$, see Eq. (\ref{newdrive}). The starting point of our approach is thus to find the solution of the expectation value $\langle\hat{a}_2\rangle$ from Eq. (\ref{exactsolution}) which, after inserting equation (\ref{newdrive}), becomes a nonlinear algebraic equation for $\langle\hat{a}_2\rangle$,
\begin{equation}
\langle\hat{a}_2\rangle=\frac{2J}{\gamma^2}\left(\Omega\left(1+e^{i\varphi}\right)-4J\langle\hat{a}_2\rangle\right)\frac{_0\mathcal{F}_2\left(-i\gamma/U,1+i\gamma/U,2\left|\frac{2iJ}{\gamma U}\left(\Omega\left(1+e^{i\varphi}\right)-4J\langle\hat{a}_2\rangle\right)\right|^2\right)}{_0\mathcal{F}_2\left(-i\gamma/U,i\gamma/U,2\left|\frac{2iJ}{\gamma U}\left(\Omega\left(1+e^{i\varphi}\right)-4J\langle\hat{a}_2\rangle\right)\right|^2\right)}.
\end{equation}
The result for $\langle\hat{a}_2\rangle$ can be put into Eq. (\ref{newdrive}) allowing for an expression of the ''new'' drive $\tilde{\Omega}$. Based on this solution the correlation values for all higher normal-ordered momenta in Eq. (\ref{exactsolution}) can be subsequently derived in a self-consistent manner.

Fig. \ref{figure22} shows the behaviour of the second order correlation function for excitations in the second cavity depending on the external experimental parameters. As expected, for low nonlinearities the state in the middle transmission line resonator is close to a coherent state. An increase of the nonlinearity leads to a state becoming more and more anti-bunched.
In contrast to the low excitation regime \cite{gerace2009quantum}, we here find that $g^{(2)}(0)$ does not monotonously approach unity as the tunneling $J$ is increased. Instead anti-bunching can also become more pronounced. We attribute this non-monotonous behavior to the competition of two effects. On the one hand, an increased tunneling rate means that more photons from the coherent fields in the outer resonators flow into the central resonator and weaken the anti-bunching. On the other hand, an increased tunneling rate detunes the normal modes of the three-cavity system from the laser drives so that the outer cavities are less populated. If this second effect starts to dominate over the previous one, the photon flow into the central resonator decreases and its excitations show more tendency to anti-bunch. This signature is also visible, albeit less pronounced, in Fig. \ref{figure19}.
\begin{figure}
\centering
\includegraphics[width=0.4\textwidth]{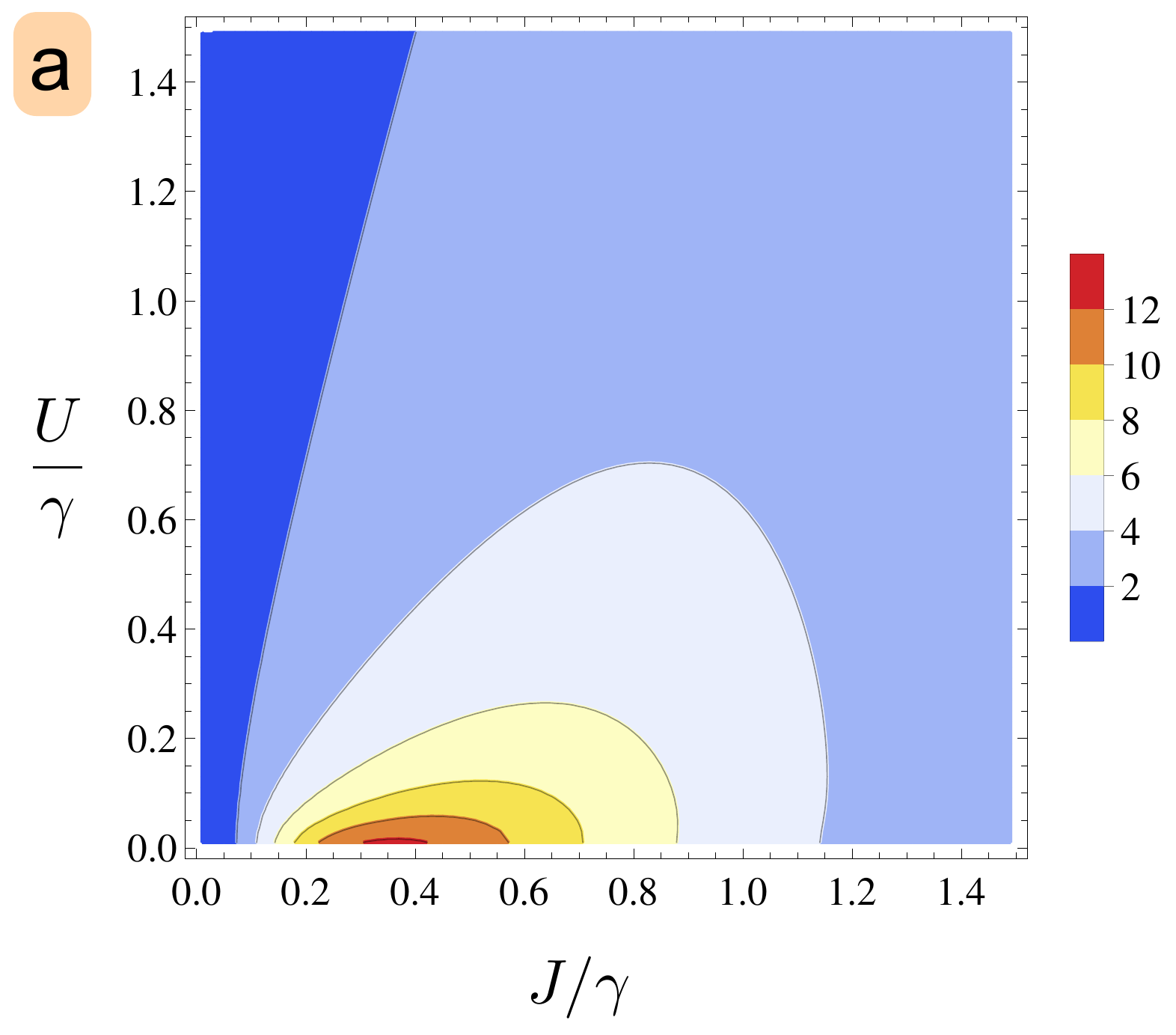}
\hspace{0.4cm}
\includegraphics[width=0.4\textwidth]{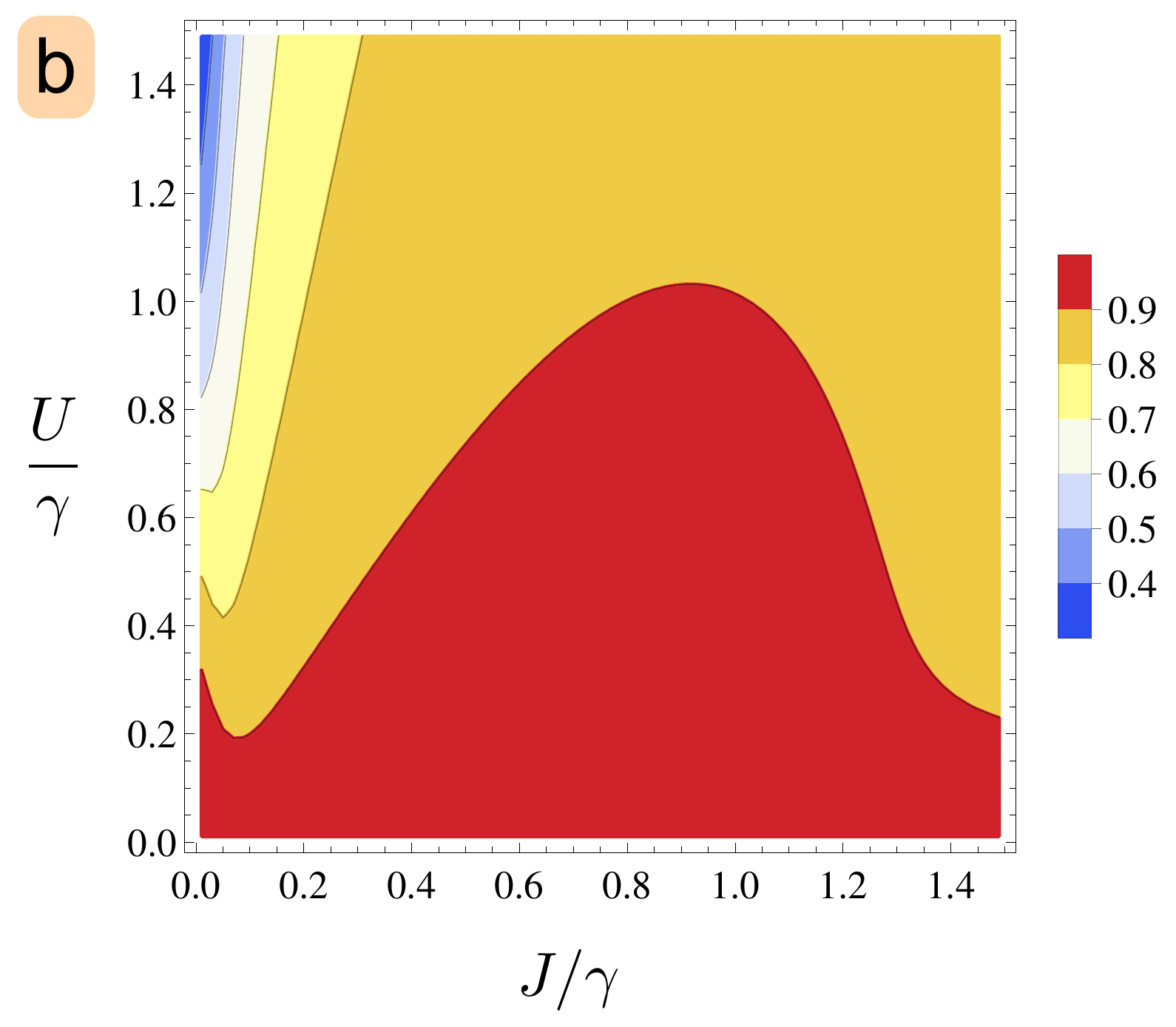}
\caption{The mean polariton number ({\bf a}) and $g^{(2)}_2$-function ({\bf b}) of the central resonator in the strong driving regime, $\Omega/\gamma=5$, as a fundtion of $J/\gamma$ and $U/\gamma$. The plots show the results gained from equation (\ref{exactsolution}). Due to the previously explained nature of our approach being a good approximation for lower tunneling strength, we are able to cover an additional regime compared to the Bogoliubov expansion, which broke down for higher values of the nonlinearity. (Compare with Fig. \ref{figure19})}
\label{figure22}
\end{figure}

In Fig. \ref{figure23} we investigate the influence of the phase-difference $\varphi$ on the excitation statistics of the central transmission line resonator. The mean excitation number fluctuates with the phase-difference $\varphi$, exhibiting maxima at $\varphi=2n\pi$ and minima at $\varphi=2(n+1)\pi$, where $n$ is an integer, due to constructive and destructive interference of the input laser drives respectively. An increase in the tunneling rate from $J/\gamma = 0.1$ to $J/\gamma = 0.5$ enables a higher number of excitations to tunnel into, and therefore excite, the central transmission line resonator, leading to a higher number of excitations altogether. The $g^{(2)}$-function also shows oscillations, however, their behavior is very different for different tunneling rates. Whereas maxima appear for both tunneling rates at a phase-difference $\varphi=2n\pi$, the second order correlation function for $J/\gamma=0.1$ shows additional local maxima at $\varphi=(2n+1)\pi$. These may be attributed to the low excitation number at these points which induce a tendency for $g^{(2)}$-functions to increase.
\begin{figure}
\centering
\includegraphics[width=0.4\textwidth]{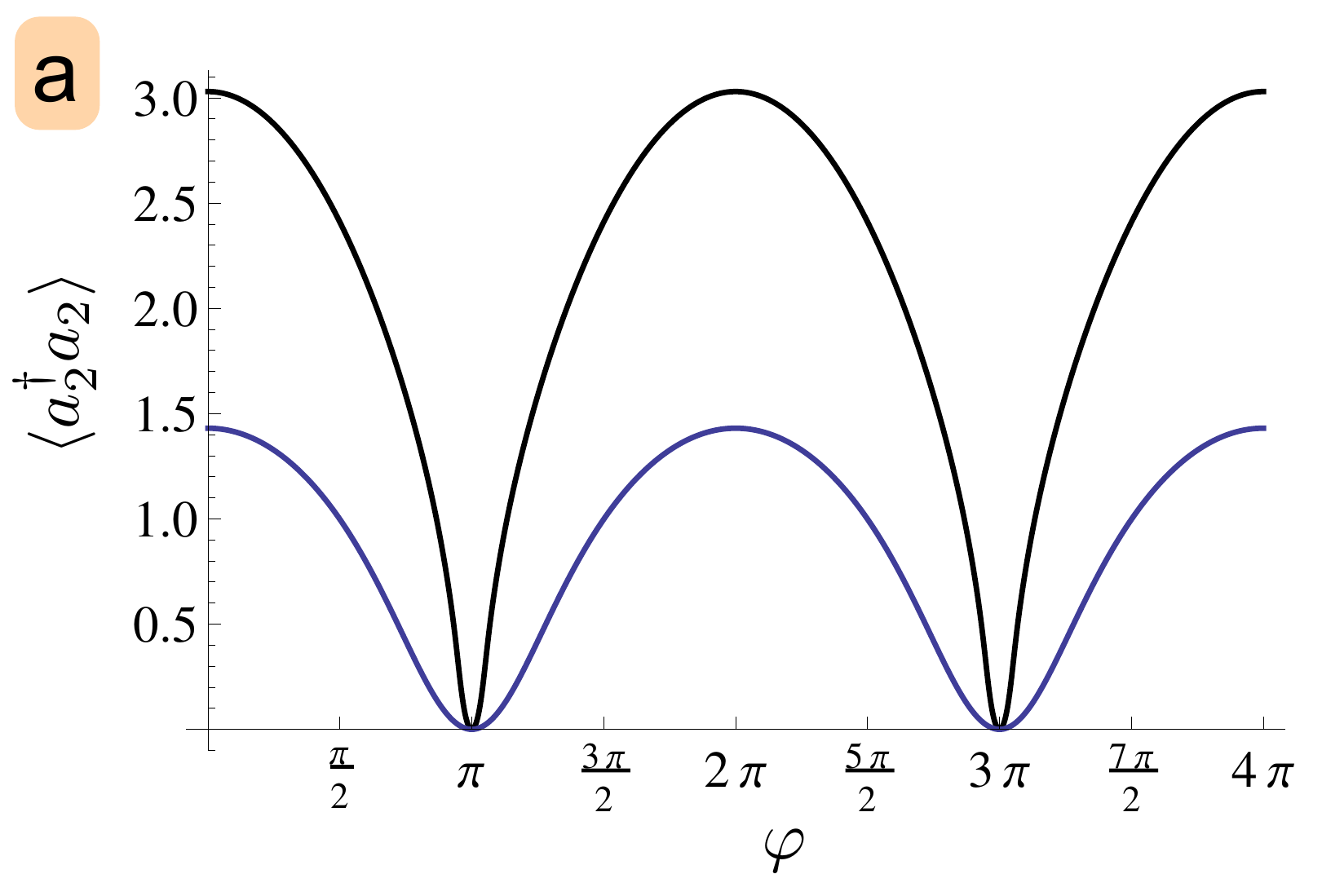}
\hspace{0.4cm}
\includegraphics[width=0.4\textwidth]{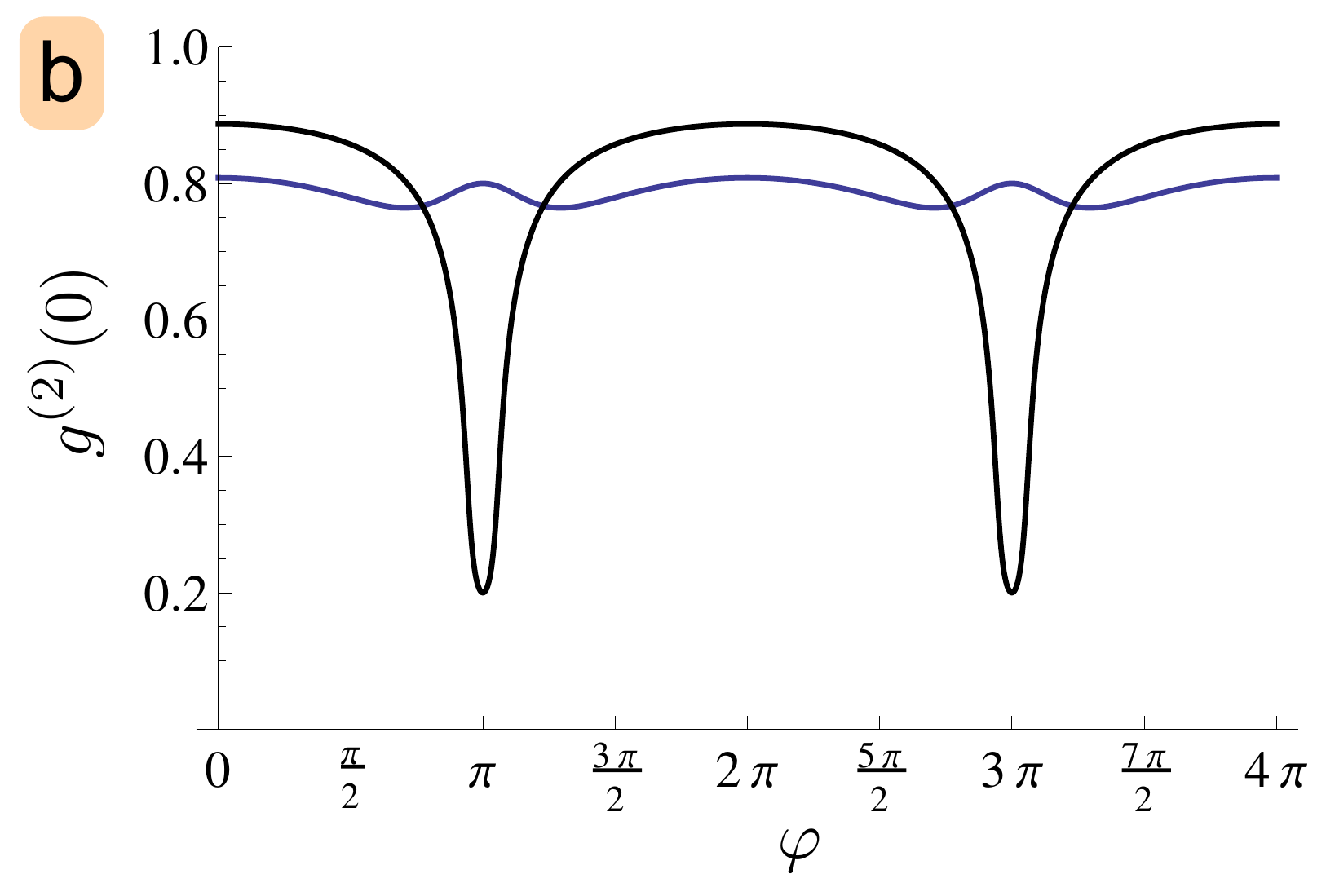}
\caption{Dependence of the second resonator's excitation statistics on the phase difference $\varphi$ of the input laser drives for $\Omega/\gamma=5$, $U/\gamma=0.5$ and $J/\gamma=0.1$ (blue) or $\Omega/\gamma=10$, $U/\gamma=2$ and $J/\gamma=0.5$ (black) as obtained from  equation (\ref{exactsolution}). {\bf a}: Mean excitation number and {\bf b}: Second order correlation function. The oscillation in the mean excitation number can be attributed to the interference of the two input laser fields.}
\label{figure23}
\end{figure}

\subsection{Comparison to the numerical solution}

Finally we quantify the parameter range where the P-function based mean-field approach we introduce here is in good agreement with a full numerical solution for regimes where the latter can be obtained. Fig. \ref{figure23a}a compares the mean-field solution (red) and full numerical solutions (dashed lines) for $\Omega/\gamma = 1.5$ and $U/\gamma = 2$ as the tunneling rate $J$ is increased. Whereas the numerical solutions for cut-offs at four (blue) and five (green) excitations agree very well indicating that the cut-offs are high enough, one can see deviations between the numerical and the mean-field solutions as $J$ increases. These should be interpreted as a failure of the mean-field approach for larger $J$.
Fig. \ref{figure23a}b in turn compares the mean-field solution (red) and full numerical solutions (dashed lines) for $J/\gamma = 0.2$ and $U/\gamma = 0.5$ as the drive strength $\Omega$ is increased. Here the numerical solutions converge to the mean-field solution as the cut-off number is increased, indicating that the mean-field approach is a very accurate approximation throughout.
The agreements and disagreements displayed in Figs. \ref{figure23a}a and b are in agreement with the intuition that enhancing the coherent drives at the outer cavities should make the fields in those cavities more coherent and hence the mean-field approximation more accurate whereas an increase in the tunneling rate enhances the influence of the nonlinearity on the outer cavities and hence drives them away from a coherent state which renders the mean-field approach inaccurate.
\begin{figure}
\centering
\includegraphics[width=0.4\textwidth]{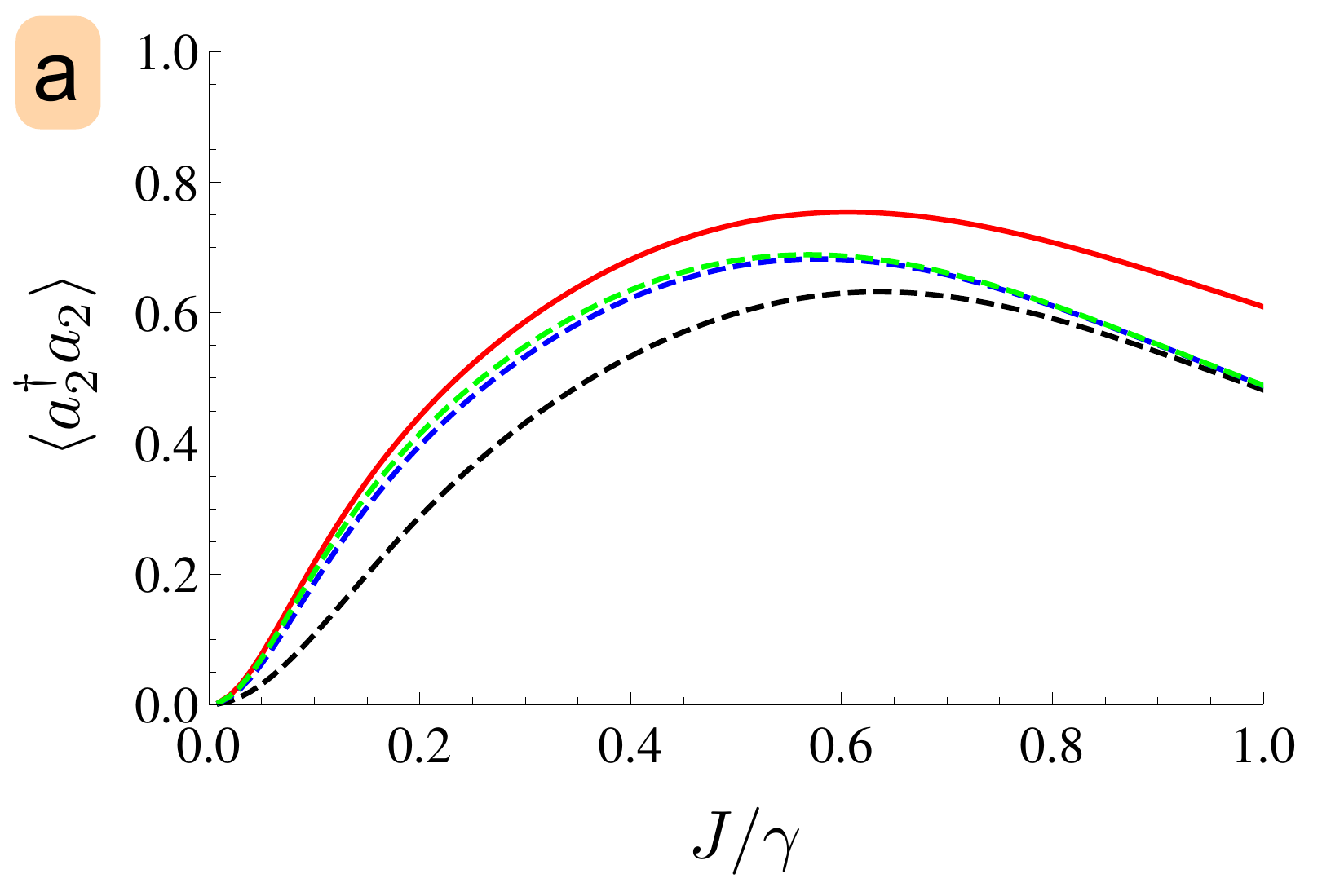}
\hspace{0.4cm}
\includegraphics[width=0.4\textwidth]{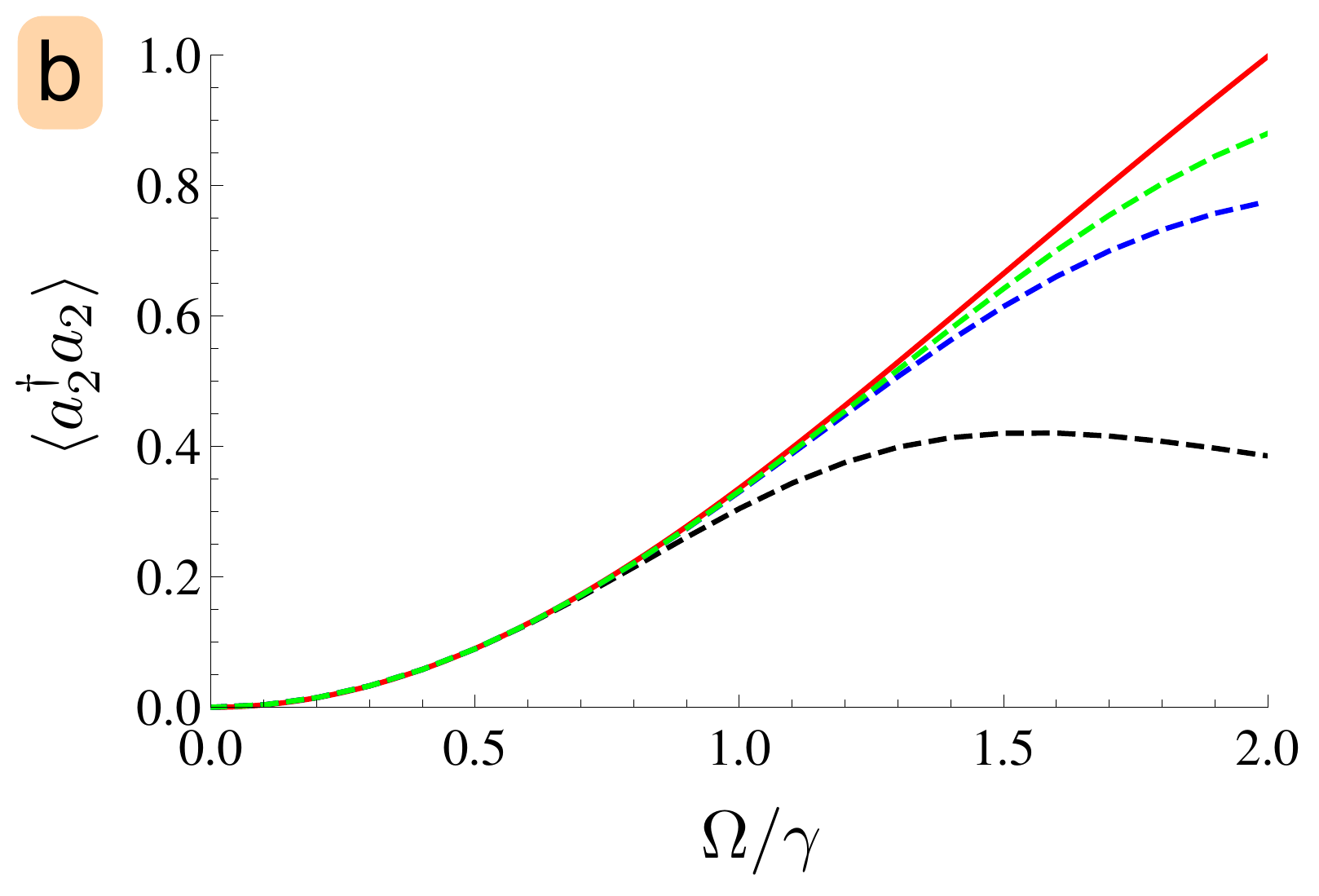}
\caption{Comparison of the results for mean polariton number in the central cavity obtained by the P-function mean-field method, red line, and by the conventional numerical method, dashed lines, with cut-offs at three (black), four (blue) and five (green) excitations. {\bf a}: Deviation due to an increase in the tunneling rate for $\Omega/\gamma = 1.5$ and $U/\gamma = 2$. {\bf b}: Deviations due to an increase in driving strength for $J/\gamma = 0.2$ and $U/\gamma = 0.5$.}
\label{figure23a}
\end{figure}

%______________________________________________________________________
%______________________________________________________________________
\section{Conclusion and Discussion of Accessible Regimes}\label{sec:conclusions}

In summary we have complemented the discussion of a quantum optical Josephson interferometer with two methods. One is a linearization or Bogoliubov expansion of the intra-cavity fields around their coherent parts. The second method, which we have introduced here, combines an exact solution for the central resonator with a mean-field decoupling of the tunneling terms to the outer resonators.

With the proposed methods the circuit QED Josephson interferometer can be described for a significant range of all experimentally adjustable parameters, such as the input drive $\Omega$, the nonlinearity $U$ and the tunneling strength $J$ for a given dissipation rate $\gamma$. If one is faced with low laser intensities and thus low intra-cavity excitations, the solution can be obtained by solving the master-equation numerically via an excitation cut-off in the matrix representation of the ladder operators as has been shown in section~\ref{numericalsection}. In a strong driving regime with intermediate to high tunneling rates but low nonlinearities a good approximation to the excitation output statistics can be given by a Bogoliubov expansion in the quantum fluctuations around the coherent background parts of the intra-cavity fields, which we presented in section~\ref{quantum fluctuations}. However, for a strong input drive, low tunneling rates and intermediate to high values of the nonlinearity the Bogoliubov expansion breaks down. Here we derived a method in section~\ref{analytical}, which we termed P-function mean-field approach, that works extremely well in said regime. The only regime for which a good approximation could not be gained was one where the drive is strong enough to produce high intra-cavity excitations, but the tunneling rate and nonlinearity are still relatively high. As a result they are responsible for non-neglectible quantum effects in comparison to the drive or to each other and therefore neither an expansion in $U$ nor in $J$ would be valid. A summary of the validity ranges for each of the presented methods is sketched in figure \ref{regions}.

%______________________________________________________________________
%______________________________________________________________________ 

%\section{New}
%
%\subsection*{Sub-heading for section}
%\subsubsection*{Sub-sub heading for section}
%\subsubsection*{Sub-sub-sub heading for section}

\bigskip

%%%%%%%%%%%%%%%%%%%%%%%%%%%%%%%%
\section*{Author's contributions}
 RJ performed the calculations and analysis, MH conceived and supervised the project, both authors discussed the results and wrote the manuscript.

%%%%%%%%%%%%%%%%%%%%%%%%%%%
\section*{Acknowledgements}
  \ifthenelse{\boolean{publ}}{\small}{}
  The authors thank Peter Degenfeld-Schonburg for fruitful discussions.
This work is part of the Emmy Noether project HA 5593/1-1 and the CRC 631, both funded by the German Research Foundation, DFG.

%%  Bmc_article.bst  will be used to                       %%
%%  create a .BBL file for submission, which includes      %%
%%  XML structured for BMC.                                %%
%%  After submission of the .TEX file,                     %%
%%  you will be prompted to submit your .BBL file.         %%

\newpage
{\ifthenelse{\boolean{publ}}{\footnotesize}{\small}
 \bibliographystyle{bmc_article}  % Style BST file
  \bibliography{Josephson_Int} }     % Bibliography file (usually '*.bib' ) 

%%%%%%%%%%%

\ifthenelse{\boolean{publ}}{\end{multicols}}{}

\end{bmcformat}
\end{document}